\documentclass[conference]{IEEEtran}
\usepackage{tikz}
\usepackage[]{graphicx}
\usepackage{subfig}
\graphicspath{{./pics/}}
\usepackage{booktabs}
\usepackage{diagbox}
\usepackage{amsmath}
\usepackage{amssymb}
\usepackage{cases}

\usepackage{bm}
\usepackage{url}
\usepackage{multirow}
\usepackage{threeparttable}
\usepackage{fancyvrb}
\usepackage{tipa}
\usepackage[normalem]{ulem}
\definecolor{casecolor}{RGB}{0,0,255}
\definecolor{casecolor2}{RGB}{255,140,0}
\newcommand{\CASE}[1]{\underline{{}#1}}
\newcommand{\CASET}[1]{\uwave{{}#1}}
\newcommand{\TN}{ACE}
\begin{document}
%----------------------------------------------------------------
\title{Compensating Removed Frequency Components: Thwarting Voice Spectrum Reduction Attacks}

\author{\IEEEauthorblockN{Shu Wang}
\IEEEauthorblockA{George Mason University\\
swang47@gmu.edu}
\and
\IEEEauthorblockN{Kun Sun}
\IEEEauthorblockA{George Mason University\\
ksun3@gmu.edu}
\and
\IEEEauthorblockN{Qi Li}
\IEEEauthorblockA{Tsinghua University\\
qli01@tsinghua.edu.cn}}

\IEEEoverridecommandlockouts
\makeatletter\def\@IEEEpubidpullup{6.5\baselineskip}\makeatother
\IEEEpubid{\parbox{\columnwidth}{
    Network and Distributed System Security (NDSS) Symposium 2024\\
    26 February - 1 March 2024, San Diego, CA, USA\\
    ISBN 1-891562-93-2\\
    https://dx.doi.org/10.14722/ndss.2024.23150\\
    www.ndss-symposium.org
}
\hspace{\columnsep}\makebox[\columnwidth]{}}

\maketitle
%----------------------------------------------------------------
\begin{abstract}
Automatic speech recognition (ASR) provides diverse audio-to-text services for humans to communicate with machines. However, recent research reveals ASR systems are vulnerable to various malicious audio attacks. In particular, by removing the non-essential frequency components, a new spectrum reduction attack can generate adversarial audios that can be perceived by humans but cannot be correctly interpreted by ASR systems. It raises a new challenge for content moderation solutions to detect harmful content in audio and video available on social media platforms. In this paper, we propose an acoustic compensation system named ACE to counter the spectrum reduction attacks over ASR systems. Our system design is based on two observations, namely, frequency component dependencies and perturbation sensitivity. First, since the Discrete Fourier Transform computation inevitably introduces spectral leakage and aliasing effects to the audio frequency spectrum, the frequency components with similar frequencies will have a high correlation. Thus, considering the intrinsic dependencies between neighboring frequency components, it is possible to recover more of the original audio by compensating for the removed components based on the remaining ones. Second, since the removed components in the spectrum reduction attacks can be regarded as an inverse of adversarial noise, the attack success rate will decrease when the adversarial audio is replayed in an over-the-air scenario. Hence, we can model the acoustic propagation process to add over-the-air perturbations into the attacked audio. We implement a prototype of ACE and the experiments show that ACE can effectively reduce up to 87.9\% of ASR inference errors caused by spectrum reduction attacks. Furthermore, by analyzing the residual errors on real audio samples, we summarize six general types of ASR inference errors and investigate the error causes and potential mitigation solutions.
\end{abstract}
% \vspace{-0.2in}
\section{Introduction}
% ASR
The development of artificial intelligence has driven the popularity of automatic speech recognition (ASR) systems, which have been integrated into human-computer interaction services such as Amazon Alexa~\cite{Alexa}, Apple Siri~\cite{Siri}, Google Assistant~\cite{GA}, and Microsoft Cortana~\cite{MicroCortana} to transform speech signals into text transcriptions. By sending voice commands, ASR systems
free users' hands and make it more convenient to control their smart home devices~\cite{alexacontrol}, send messages to their friends~\cite{CarrascoMolina2019}, or navigate to their destinations~\cite{drivernavigate}. 

% ASR Attack
% \vspace{-0.02in}
Meanwhile, recent studies have shown ASR systems are vulnerable to various malicious voice attacks~\cite{Impersonation, ModIndex, cm_1, Lav2017, modreplay, InaudibleVC, Dolphin, inaudible_0, BackDoor, SPHidden, AdversarySamples, Psychoacoustic, Dompteur, HidCmd, CmdSong, SpecPatch}.
% An ASR system typically contains four stages: audio capture, signal preprocessing, feature extraction, and model inference~\cite{SPHidden}.
%
%Particularly, in the audio capture stage, the attackers can launch  impersonation attacks~\cite{Impersonation, ModIndex}, audio replay attacks~\cite{cm_1, Lav2017, modreplay}, or inaudible audio attacks~\cite{InaudibleVC, Dolphin, inaudible_0, BackDoor}. 
%Attackers can also leverage the malicious audio conversion to attack ASR in the signal preprocessing and feature extraction phases~\cite{SPHidden}.
%Furthermore, by exploiting the vulnerabilities of deep learning, adversarial audio attacks can manipulate the inference results of the speech recognition models~\cite{AdversarySamples, Psychoacoustic, Dompteur, HidCmd, CmdSong, SpecPatch, DevilWhisper, AdvPulse}. 
% For instance, attackers can generate an adversarial audio sample, which can only be recognized by ASR but cannot be interpreted by humans.
% DFT attack
As an alternative representation of voice signals, frequency spectrum has been manipulated by attackers to achieve different attacking goals. By adding high frequency components out of the voice band, attackers can launch spectrum addition attacks to generate the audio that can be interpreted by machines but incomprehensible to humans~\cite{SPHidden}. In contrast, by removing the frequency components of weak strength from the audio spectrum, an attacker can launch spectrum reduction attacks to generate the audio that can be perceived by humans but cannot be correctly interpreted by machines~\cite{dft_attack, abdullah2022attacks}.
Moreover, attackers can manipulate the spectrum magnitude with a specific filter to bypass the spectrum-based detection mechanisms~\cite{modreplay}.
The spectrum addition attacks can be effectively mitigated by voice band-pass filters~\cite{SPHidden}; however, there is no effective defense mechanism resolving the spectrum reduction attacks.

% Recently, researchers proposed multiple voice attack methods by modifying the voice spectrum~\cite{dft_attack, SPHidden, modreplay}.
% Especially, a new Discrete Fourier Transform (DFT) attack can generate audios that can be perceived by humans but cannot be correctly interpreted by the ASR systems~\cite{dft_attack}.

\vspace{-0.015in}
Spectrum reduction attack removes certain frequency components whose magnitudes are less than a specific threshold. Though the removed frequency components have low intensity, they are critical for the correctness of ASR interpretation. In other words, the component removal may change the transcription results of ASR systems by altering spectrum distribution~\cite{davis1980comparison}. Meanwhile, since these components are non-essential for human comprehension, humans can still understand the audio even with the component removal. It becomes a new challenge to the social media platforms, which usually pre-screen and filter out harmful content such as misinformation and violence with their content moderation systems.
With the voice spectrum reduction attacks, malicious influencers can post and spread the videos or audios containing restricted speeches to online users without triggering any content alerts. This is because the sensitive content within the audio tracks cannot be noticed or detected by machine-based detection systems; meanwhile, the harmful information can be perceived by humans and thus have a negative influence on public audiences.
% The hypothesis of the DFT attack is that ASR systems rely on some weak frequency components that are non-essential for human comprehension.
%The DFT attack may induce ASR systems to execute undesired commands without the notice of human beings.
Moreover, the spectrum reduction attack can be easily launched without using any specific devices (e.g., ultrasonic generator or laser emitter) or any deep learning knowledge, so it is generic and practical to any ASR systems.
{For example, researchers have shown this attack is effective on Google Speech-to-Text APIs, Facebook Wit, Deep Speech, GMU Sphinx, and Microsoft Azure Speech APIs~\cite{dft_attack}.}

% is not specific to any ASR systems.

\vspace{-0.015in}
% our work, motivation, (1) frequency compensation (2) acoustic propagation emulation
In this paper, we present an acoustic compensation system named \TN{} to counter the spectrum reduction attacks.
The basic idea is to recover the removed frequency components according to the remaining spectrum and to enhance the audio robustness by introducing appropriate perturbations.
% \TN{} is designed based on two defense strategies: frequency compensation strategy and acoustic propagation strategy.
Our design is based on two observations: {\em frequency component dependencies} and {\em perturbation sensitivity}. First, the Discrete Fourier Transform (DFT) computation involved in the attacks can introduce spectral leakage effect and sometimes aliasing effect to the audio frequency spectrum, inevitably leading to a high correlation among the frequency components with similar frequencies.
Due to the intrinsic dependencies between neighboring frequency components, it is possible to recover more of the original audio by compensating for the removed components according to the remaining ones. Second, the spectrum reduction attack could be considered as an adversarial attack, where the removed components are an inverse of added adversarial noise. Due to the perturbation sensitivity of adversarial noise, the attack success rate would decrease when the audio is played in an over-the-air scenario~\cite{li2019adversarial}. Thus, to mitigate the spectrum reduction attacks, we introduce the over-the-air perturbations into attacked audio by modeling the acoustic propagation process (e.g., the ambient noise).
% and multipath effect

% \vspace{-0.02in}
% design and implementation.
% 4 modules, compensation module, noising module, echo module, adaptive module.
The \TN{} system contains three main modules, including two core functional modules (i.e., spectrum compensation module, noise addition module) connected in sequential order for audio signal conversion and an auxiliary module (i.e., adaptation module) for attack detection and parameter estimation.

\vspace{-0.02in}
\noindent {\bf Spectrum Compensation Module.}
The spectrum compensation module estimates the original audio spectrum by aggregating the scaled shifted attacked spectrum. The compensation operation is mathematically equivalent to the linear convolution of the remaining frequency spectrum and a set of scaling coefficients. Therefore, we can obtain the filter parameters of the spectrum compensation module by linear regression.

\vspace{-0.02in}
\noindent {\bf Noise Addition Module.}
The noise addition module emulates the Gaussian white noise during the audio propagation, which is equivalent to adding equal intensity to all frequency components. Since the magnitudes of removed components are under a threshold, the added noise with limited magnitude can be seen as an approximate alternative to the missing components. Meanwhile, since the noise is too weak to affect the remaining strong components, it can help recover the original spectrum distribution without changing the human comprehension.
% 
% The echo module aims to emulate the multipath effect caused by audio reflections. To emulate the perturbations, the echo module generates the received signal by shifting and scaling the audio signals in the time domain and aggregating all delayed signals.

\vspace{-0.02in}
\noindent {\bf Adaptation Module.}
To achieve the highest ASR inference accuracy under any given situation, the adaptation module is responsible for estimating the component removal ratio by calculating the proportion of extremely weak components and configuring the optimal parameters for two core modules. When the attackers are aware of our defense, they may launch an adaptive attack that changes component removal ratios to circumvent our defense. In this case, the adaptation module is responsible for quickly  detecting the attack parameter and updating the module parameters for the current audio segments.

% \vspace{-0.02in}
% exp results.
% we did extensive exp, DFT attack, evaluation metric, (word-err, character-err). 
We implement a prototype of \TN{} and study its performance by conducting extensive experiments over two popular ASR systems (i.e., DeepSpeech~\cite{deepspeech} and CMU Sphinx~\cite{CMUSphinx}) and two public audio datasets {TIMIT}~\cite{TIMIT} and VCTK~\cite{VCTK}. We first reproduce the spectrum reduction attacks in two settings: phoneme-level attacks and word-level attacks.
To evaluate the mitigation performance of \TN{}, we measure the ASR inference differences between the attacked audio and our mitigated audio by using the word error rate (WER) and the character error rate (CER), which reflect the minimum word-level and character-level edit distance towards the real transcriptions~\cite{wer}.
Under the spectrum reduction attacks with a relatively high component removal ratio (e.g., 85\%), the \TN{} can reduce the average WER from 0.90 to 0.57 and reduce the average CER from 0.71 to 0.42. Considering even the benign audio retains an inherent WER of 0.49 and a CER of 0.38, \TN{} can effectively eliminate up to 79.5\% WER and up to 87.9\% CER among the mitigable errors (i.e., the added errors actually caused by attacks). % within the feasible extent.
% the compensation module can reduce WER from 0.60 to 0.33 (43.7\% decrease) and reduce CER from 0.38 to 0.20 (47.5\% decrease).
% With the noising module, the WER (CER) can be reduced by up to 46.0\% (50.9\%).
% Also, the WER (CER) will decrease by up to 11.8\% (14.0\%) if we use the echo module.
We also evaluate the individual performance of each module and analyze the parameter selection under different scenarios to achieve the optimal performance. %We find both the spectrum compensation module and noise addition module contribute largely to the~\TN{}.
In addition, even against the most aggressive adaptive attacks that update the parameters every 80 ms (i.e., the phoneme duration), \TN{} can still mitigate the recognition errors by 82.3\% using the adaptation module.
% with the highest parameter update frequency of every 80 ms, 
%, while the echo module also plays an important role in the error mitigation. 
The experimental results show \TN{} can efficiently reduce the ASR inference errors caused by spectrum reduction attacks, thus improving the ASR robustness.

% \vspace{-0.02in}
% case study on the ASR transcription of .., .., and analyze the error provenance.
To further analyze the root cause of ASR inference errors and better understand why the \TN{} can mitigate these errors, we perform an error analysis by looking into the audio examples that result in the wrong transcriptions.
We find that ASR inference errors can be grouped into six categories, namely, elision errors, rare word errors, consonant errors, vowel errors, shifted phoneme errors, and natural language processing (NLP) inference errors.
For benign audio, the ASR inference errors are mainly attributed to the rare words (40.3\%) and elision (20.9\%), which rely on the corpus and language model selection.
However, under spectrum reduction attacks, the inference errors mainly occur in phonetic misinterpretation. For instance, 73.7\% samples contain consonant errors, 68.4\% samples contain vowel errors, and 47.4\% samples contain shifted phoneme errors.
By applying our \TN{} system, the vowel errors have been largely diminished, only remaining in 22.7\% samples.
Due to the higher loudness and signal strength, vowels carry richer information for restoring the phonetic spectrum distribution.
However, we also find the remaining errors are concentrated in consonants (63.6\%) since the weaker strength and shorter duration make them more likely to be confused.
% they are more likely to be confused due to the lighter strength and shorter duration.
% focus on the consonant errors (63.6\%) while the 
%
Therefore, we conclude that the spectrum reduction attacks affect the ASR inference results mainly via phonetic errors and the \TN{} system can largely mitigate the inference errors by correcting the vowels.

% \vspace{-0.01in}
% contributions.
In summary, our paper makes the following contributions:

% \vspace{-0.01in}
\begin{itemize}
    \item We propose an acoustic compensation system called \TN{}, which mitigates the spectrum reduction attacks by performing frequency spectrum compensation and introducing additional acoustic perturbations. 
    % \vspace{-0.01in}
    \item We implement a prototype of the \TN{} system with spectrum compensation, noise addition, and adaptation modules to process the incoming audio according to the estimated attack settings.
    % \vspace{-0.01in}
    \item We conduct extensive experiments to show that \TN{} can adaptively reduce the ASR inference errors caused by spectrum reduction attacks with different attack granularity.
    % \vspace{-0.01in}
    \item By performing a residual error analysis on real samples, we summarize 6 types of ASR inference errors and analyze their root causes along with potential error mitigation methods. 
\end{itemize}
% \vspace{-0.1in}
% \vspace{-0.1in}
\section{Preliminaries}

% We first introduce the background knowledge about automatic speech recognition, audio signal processing, and DFT attack.

% \vspace{-0.05in}
\subsection{Automatic Speech Recognition}

% \begin{figure}[t]
%     \centering
%     \includegraphics[width=\linewidth]{pics/ASR.pdf}
%     \caption{The pipeline of automatic speech recognition.}
%     \label{fig:asr_pipeline}
% \end{figure}

% audio capture, signal pre-processing, feature extraction, model inference

Automatic speech recognition (ASR) provides the audio-to-text conversion with four steps, namely, audio capture, signal pre-processing, feature extraction, and model inference~\cite{SPHidden}. First, a microphone captures audio signals by converting air vibrations into electronic signals, which are then transformed into a digital audio format via an analogy to digital converter. Second, the digital audio signals are preprocessed by denoising algorithms and a low-pass filter with a typical cut-off frequency of 8 kHz. Signal preprocessing phase is critical for improving ASR recognition accuracy. Third, the processed audio is split into overlapping frames, which are used to extract the audio features. The most common feature is the Mel Frequency Cepstral Coefficient (MFCC)~\cite{MFCCfeature}. To obtain MFCC, Discrete Fourier Transform (DFT)~\cite{DFT} is used to first get the audio spectrum, where the frequencies are converted into the Mel scale and the magnitudes are converted into the logarithm scale. Then, MFCC can be calculated by taking the Discrete Cosine Transform (DCT) of the list of Mel filter bank energies. Finally, the audio features are fed into a deep learning model to get the inference results, i.e., the content of the input speech.

% \vspace{-0.03in}
\subsection{Audio Signal Processing}

Typically, audio signals can be analyzed from two different perspectives, i.e.,  time domain and frequency domain. In the time domain, an audio signal is represented as $s(t)$, which records the relative signal intensity at each instant. The same signal would also be represented in the frequency domain, i.e., $S(f)$, which reflects multiple sinusoidal components at each frequency. Fourier transform is the bridge to achieve the time-frequency domain conversion. Since modern devices process the audio signals in digital format, the $N$-point Discrete Fourier Transform (DFT) is utilized to convert the time-domain signal $s(n)$ into a frequency-domain signal~\cite{DFT}.
\begin{equation}
    S(k) = \sum_{n=0}^{N-1} s(n) \cdot [cos(\frac{2 \pi}{N} kn) - i \cdot sin (\frac{2 \pi}{N} kn)],
\end{equation}
\noindent where $S(k)$ is a sinusoidal frequency component with the frequency of $f_{s} \cdot k/N$ ($f_s$ is the sampling rate). The absolute value $|S(k)|$ indicates the magnitude of the component and the argument $arg[S(k)]$ indicates the phase of the component. All the frequency components $\{S(k), k \in [0, N-1]\}$ compose the entire signal spectrum. To convert the frequency-domain signal back to time domain, we can use inverse DFT (IDFT).

% \vspace{-0.03in}
\subsection{Spectrum Reduction Attack}

Voice spectrum reduction attack is based on the hypothesis that ASR systems rely on some frequency components that are non-essential for human comprehension~\cite{dft_attack}.
Thus, even if these components are removed, human listeners can still recognize the modified audio; however, ASR systems may misinterpret the audio and output wrong transcriptions. To launch spectrum reduction attack, the weak frequency components under a specific threshold are removed from the original audio signals.

\begin{figure}[t]
    \centering
    % \vspace{-0.03in}
    \includegraphics[width=3.3in]{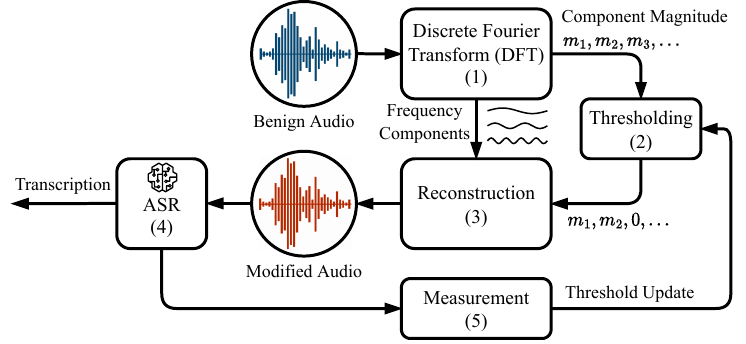}
    % \vspace{0.1in}
    \caption{Workflow of Spectrum Reduction Attack.}
    \label{fig:dft_attack}
    \vspace{-0.2in}
\end{figure}

Figure~\ref{fig:dft_attack} shows the workflow to generate malicious audio via the spectrum reduction attack. First, given a benign audio, attackers first segment the audio signal and utilize DFT to decompose each audio segment into multiple frequency components. Second, the attackers remove the frequency components whose magnitudes are lower than a threshold. Third, the remaining frequency components are reconstructed back to a time-domain audio by IDFT. Fourth, the ASR system provides the transcription for the modified audio. Fifth, the transcription differences between the inference and the ground truth are measured to update the threshold, e.g., enlarging the threshold for too small differences. The threshold is determined until the differences no longer change and humans cannot notice the perception differences. Here, the proportion of removed frequency components is called \emph{component removal ratio}, which is the only parameter in the spectrum reduction attack. According to the granularity of signal segments, the attacks can be launched at word level or phoneme level, where the word-level attacks perform spectrum reduction on each word and the phoneme-level attacks take phonemes as basic units.

\begin{figure*}[t]
    \centering
    \includegraphics[width=0.98\linewidth]{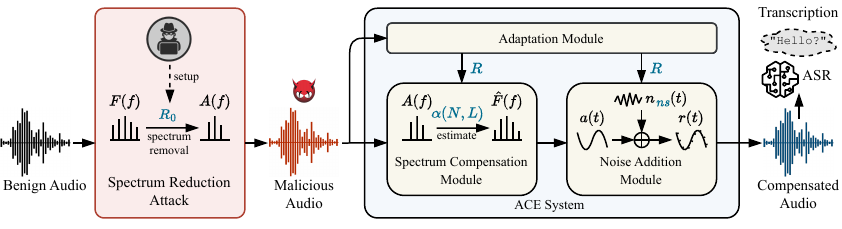}
    % \vspace{-0.08in}
    \caption{The overview of acoustic compensation system (\TN{}) against spectrum reduction attack.}
    \label{fig:overview}
    \vspace{-0.2in}
\end{figure*}

% \vspace{-0.20in}
\section{Threat Model}
% attack
Attackers can leverage either phoneme-level or word-level spectrum reduction attacks to convert benign audio into adversarial audio using a specific component removal ratio ($R_0$). %Though human listeners can recognize the audio content, ASR systems may misunderstand the malicious audio due to the changes of spectrum distribution. 
%Spectrum reduction attacks aim at generating the audio that can be perceived by humans but cannot be correctly interpreted by machine.
We assume attackers can access the original audio and edit the audio via  digital signal processing (e.g., DFT) at either software or hardware level. Attackers know real speech content and utilize a specific value of $R_0$ to remove the frequency components of weak strength from the original audio spectrum, misleading ASR systems for wrong inference results.

% defender
% As a defense system against spectrum reduction attacks, 
We assume the \TN{} system has permission to access and transform the audio prior to the ASR APIs. It is unnecessary for defenders to know the speech content or the component removal ratio used by attackers. The defenders only know the basic attack principle. The \TN{} implementation includes training and testing phases. In the training phase, \TN{} can access both the original and attacked spectrums to learn the fitting parameters in each module. However, in the testing phase, \TN{} only needs to access the attacked spectrum and cannot access the original audio (i.e., ground truth).
After the processing of \TN{}, we assume the pipeline between \TN{} and ASR is secure. 
We trust the ASR system and assume attackers cannot modify acoustic/language models and vocabulary files.

%adaptive attack
If the attackers are aware of \TN{} defense system, they may launch adaptive attacks that manage to use time-varying component removal ratios to circumvent our defense and achieve the original attacking goal, namely, generating audio that can be perceived by humans but cannot be correctly interpreted by machines. Besides, attackers may launch attack variants by attenuating weak component magnitude instead of removing them or removing frequencies over specific frequency bands.

\vspace{-0.02in}
\section{System Design}
\vspace{-0.05in}

\subsection{Overview}
%
%For audio adversarial attacks, attackers add adversarial noise to original audio signals so that machines would misunderstand the audio while humans can still recognize them~\cite{adversarialML}. 
We design an acoustic compensation system named \TN{} to mitigate the effects of spectrum reduction attacks. The design of \TN{} is derived from two observations. First, the adjacent frequency components of natural sound have a high correlation, which makes it possible to estimate the removed components according to the remaining neighboring ones. Thus, we utilize a frequency compensation strategy to recover the original audio. Second, the spectrum reduction attack is one type of adversarial attacks, which are sensitive to small perturbations in both the frequency and time domains. When adversarial audio is played in an over-the-air environment, the attack success rate would decrease dramatically due to the inevitable perturbations~\cite{KWS}. Hence, we emulate the over-the-air perturbations into the malicious audio to mitigate the attack effects by utilizing an acoustic propagation strategy.
% . Therefore, we design an acoustic propagation strategy to emulate the noise in real scenarios.

% \vspace{-0.05in}

Figure~\ref{fig:overview} shows the overview of \TN{} that contains three main modules, namely, spectrum compensation module, noise addition module, and adaptation module. The spectrum compensation module is based on the frequency compensation strategy that aims to reconstruct the removed weak components using the remaining strong ones. In Figure~\ref{fig:overview}, \TN{} estimates the original audio by the filter parameter $\alpha$, while the hyper-parameters $N$ and $L$ are spectrum and segment lengths, respectively. The noise addition module is based on the acoustic propagation strategy to introduce the over-the-air perturbations (i.e., noise with strength of $ns$) into input audio. The adaptation module focuses on detecting the attack and estimating the real spectrum removal ratio $R$, which is then used to determine the optimal parameters of $\alpha$ and $ns$ for spectrum compensation and noise addition modules. The goal of the adaptation module is to always achieve the highest ASR inference accuracy under any attack parameter even if the attacker is aware of the deployment and internal design of our defense system.

% \vspace{-0.03in}
\subsection{Spectrum Compensation Module}
% \vspace{-0.03in}
% \subsubsection{Frequency Compensation Strategy}

% reason
% DFT attack involves the basic signal processing operation - Discrete Fourier Transform.
% However, the DFT computation could introduce aliasing, spectral leakage, and picket fence effect to the frequency spectrum.
% All of these spectrum effects enhance the intrinsic dependency between the neighboring frequency components.
% Based on the correlations between frequency components, it is possible to estimate and recover the deleted weak components according to the remaining strong ones.
% Hence, we propose the frequency compensation strategy to restore the spectrum in the frequency domain.

% solution
% We achieve the frequency compensation strategy by the compensation module.
% In the compensation module, we compensate for the missing frequency components by shifting and scaling the neighboring strong components. 
% This compensation operation is mathematically equivalent to the linear combination between the remaining frequency components and a set of scaling coefficients.
% Therefore, by fitting multiple speech samples from public datasets, we can obtain the coefficients using linear regression. 

%%%%%

\begin{figure}[h]
    \centering
    % \vspace{-0.05in}
    \includegraphics[width=0.85\linewidth, height=1.87in]{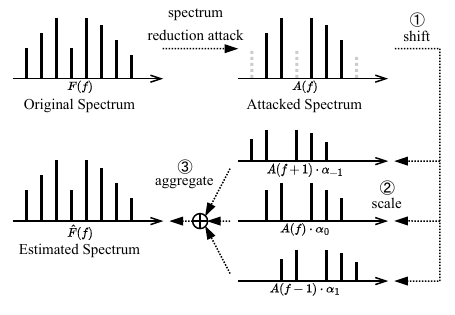}
    \vspace{-0.1in}
    \caption{The working flow of spectrum compensation module. The attacked spectrum is first shifted and scaled, then all  sub-spectra are aggregated to estimate the original spectrum.}
    \label{fig:comp_module}
    \vspace{-0.17in}
\end{figure}

%% General intro.
To defeat spectrum reduction attacks, an intuitive idea is to recover the deleted components based on the existing ones. 
In Figure~\ref{fig:comp_module}, we design a spectrum compensation module to restore the original audio frequency spectrum by shifting, scaling, and aggregating the attacked frequency spectrum.

% , by which we tend to restore the original frequency spectrum by shifting, scaling, and aggregating the mutilated frequency spectrum, as shown in Figure~\ref{fig:comp_module}.

% 
%
% These deleted components are non-essential for human comprehension but are the supporting factors to the voice processing systems.
%
% In the compensation module, we tend to restore the original frequency spectrum by shifting, scaling, and aggregating the mutilated frequency spectrum, as shown in Figure~\ref{fig:comp_module}.

%\vspace{0.03in}
\noindent{\bf Hypothesis.}
Our hypothesis is the adjacent frequency components have a high correlation, which is derived from two side effects introduced by the DFT computation, namely, (i) \emph{spectral leakage} caused by signal truncation and (ii) \emph{aliasing} caused by signal under-sampling. The spectral leakage effect introduces derived correlations to the neighboring frequency components over the computed DFT spectrum. The aliasing effect, which only occurs in devices with low sampling rates, can lead to the overlapping of frequency components. Due to these effects, the neighboring components are no longer independent and the amplitude changes along frequency components would not be abrupt. Thus, the intrinsic dependency among frequency components is the basis of spectrum compensation model. Based on this hypothesis, we are able to estimate and restore the removed frequency components using the remaining ones.

% It is derived from three side effects introduced by the DFT computation, namely, (i) \emph{aliasing} caused by signal sampling, 
% % if freq_sig > freq_samp/2, aliasing
% (ii) \emph{spectral leakage} caused by signal truncation, and 
% % signal truncation -> add t-window -> generate other components.
% (iii) \emph{picket fence effect} caused by DFT sampling.
% % the discrete sampling -> frequency is not accurate.
% % 
% Due to these side effects, the neighboring components are no longer independent over the calculated DFT spectrum. 
% Therefore, the amplitude changes would not be abrupt along the frequency components. 
% The intrinsic dependency between the neighboring frequency components is the foundation of the spectrum compensation model.
%
% Based on this hypothesis, we aim to predict and restore the missing frequency components according to the remaining components with higher intensity.

% \vspace{0.03in}
\noindent{\bf Modeling.}
Our method first shifts the attacked spectrum $A(f)$ by $i\ (-L \leq i \leq L)$ DFT units, where $L$ is the maximum shifting unit that indicates the window size of prediction. Then, the shifted spectrum $A(f-i)$ will be scaled with a scaling factor $\alpha_{i}$. Finally, all the scaled shifted spectra are aggregated as $\hat{F}(f)$ to estimate the original spectrum $F(f)$.
% \vspace{-0.05in}
\begin{equation}
    \hat{F}(f) = \sum_{-L \leq i \leq L} \alpha_{i} \cdot A(f-i),
% \vspace{-0.05in}
\end{equation}
\noindent where $\alpha_{i}\ (-n \leq i \leq n)$ are the parameters of the inverse filter that converts attacked spectrum back to the original one.

% \vspace{-0.1in}
\noindent{\bf Parameter Fitting.}
In the training phase, we obtain the scaling factors $\alpha_{i}\ (-n \leq i \leq n)$ by fitting the original and attacked spectrums. For an original spectrum $F(f)$, we can generate the attacked spectrum $A(f)$ and then utilize linear regression to obtain $\alpha$. To obtain the filter parameter $\alpha = [\alpha_{-L},\ \alpha_{-L+1},\ ..,\ \alpha_{L-1},\ \alpha_{L}]^{T}$, we first construct the Hanker matrix $H_{N \times (2L+1)}$ and a spectrum vector $F_{N \times 1}$, where $H \cdot \alpha = F$ and $N$ is the total spectrum length.
% \vspace{-0.05in}
\begin{equation}
\resizebox{.9\hsize}{!}{$
\left[
\begin{matrix} 
A(-L) & A(-L+1) & .. & A(L-1) & A(L) \\
A(-L+1) & A(-L+2) & .. & A(L) & A(L+1) \\
.. & .. & .. & .. & ..\\
A(-L+N-2) & A(-L+N-1) & .. & A(L+N-3) & A(L+N-2) \\
A(-L+N-1) & A(-L+N) & .. & A(L+N-2) & A(L+N-1) \\
\end{matrix}
\right]
\cdot 
\left[
\begin{matrix} 
\alpha_{-L}  \\
\alpha_{-L+1}   \\
..\\
\alpha_{L-1}  \\
\alpha_{L}  \\
\end{matrix}
\right]
=
\left[
\begin{matrix} 
F(0)  \\
F(1)  \\
..\\
F(N-2)  \\
F(N-1)  \\
\end{matrix}
\right]
$}
% \vspace{-0.05in}
\end{equation}
Since the spectrum length $N$ is larger than the filter size $(2L+1)$, $\alpha$ can be calculated as 
\begin{equation}
\label{eq:lr}
    \alpha = (H^{T} \cdot H)^{-1} \cdot H^{T} \cdot F.
% \vspace{-0.05in}
\end{equation}
After obtaining the filter parameter $\alpha$, we only access the attacked audio in the testing phase. We estimate the original spectrum using $\hat{F}(f) = A(f) * \alpha$, where $*$ means convolution.

% \vspace{0.03in}
\noindent{\bf Adaptive Compensation.}
Since attackers may adopt different attack settings (i.e., different component removal ratios), a spectrum compensation module with a fixed set of parameters might not always perform well under different scenarios. To adapt to malicious audio with different settings,  the spectrum compensation module includes an adaptation mechanism. First, we train multiple models using the original audio and the attacked audio with different component removal ratios. Thus, each model has the best performance against a specific component removal ratio. After estimating the component removal ratio, we use the model of the closest value to compensate for the spectrum distortion caused by spectrum reduction attacks.

%=================================================================

% \vspace{-0.03in}
\subsection{Noise Addition Module}

The noise addition module emulates and introduces the noise perturbations into audio signals to improve audio robustness, based on the fact that the over-the-air perturbations can reduce the attack success rate of adversarial audio signals.

% Based on the fact that the over-the-air perturbations can reduce the attack success rate of adversarial voice signals, we propose the acoustic propagation strategy.
% We emulate the perturbations in the sound propagation channel and bring the perturbations into the audio signals to improve their robustness.
% We emulate two main effects during the acoustic propagation, i.e., the ambient noise and the multipath effect.

% noising
% We emulate the ambient noise by the noising module, which adds Gaussian noise of limited strength to the input audio signals in the time domain.
% From the perspective of frequency domain, the added Gaussian noise has equal intensity at different frequencies.
% Because the deleted frequency components are weak under a threshold, the Gaussian noise can be seen as an approximate replacement of missing components.

% Meanwhile, the added noise will not affect the strong frequency components due to the limited strength.

% By adding the emulated ambient noise, we can recover the spectrum distribution and increase speech recognition accuracy.

%%%

% Adding noise is also an effective way to mitigate the voice attacks since the attacked voice signals are sensitive to small artificial perturbations.
% Therefore, in the noising module, we tend to add a little Gaussian noise to the attacked signals in the time domain so that the modified signals can be inferred successfully. 

\begin{figure}[t]
    \centering
    \includegraphics[width=0.99\linewidth]{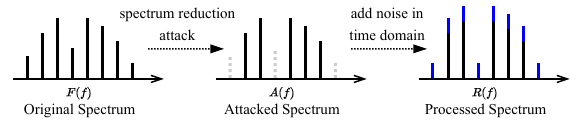}
    % \vspace{-0.08in}
    \caption{In the noise addition module, the added noise can fill in the removed weak components while the distribution of the processed spectrum can be similar to that of the original one.}
    \label{fig:noising_module}
    \vspace{-0.2in}
\end{figure}

% \vspace{0.03in}
{\noindent \bf Hypothesis.}
The principle of the noise addition module is based on the statistical properties of removed frequency components. We can consider the removed components as an added special adversarial noise, whose effect is to counteract (remove) the weak components in the frequency domain. If $S_{f}$ denotes the frequency set of all removed components, the equivalent adversarial noise can be represented as follows.
% \vspace{-0.02in}
\begin{equation}
    n_{adv}(f) = -\sum_{f \in S_{f}} | m_{f} \cdot e^{j  (2 \pi f +\phi_{f})} |,
% \vspace{-0.05in}
\end{equation}
\noindent where $m_{f}$ and $\phi_{f}$ are the magnitude and phase of the removed component of frequency $f$. We can find $n_{adv}(f)$ has a similar property with the Gaussian noise of a limited intensity since (1) all magnitude $m_{f}$ of $n_{adv}(f)$ are under a small threshold and (2) Gaussian noise has an equal magnitude at each frequency. Therefore, if the noise intensity is set to the same threshold, adding Gaussian noise can be seen as an approximate recovery of removed weak components, as shown in Figure~\ref{fig:noising_module}. However, adding noise also increases the magnitude of remaining strong components. Fortunately, the effect of noise addition on strong components is limited since the added magnitude is not comparable to the original one. Hence, if we control the noise magnitude carefully, the noise addition module can recover part of the original inference results from the attacked audio.

% discuss the effect.
Note that the removed weak components would not affect human comprehension because they carry less information. Similarly, human listeners cannot notice the added noise with a limited magnitude. The voice inference of ASR systems is derived from statistical features of audio frequency spectrum. For example, the most common acoustic feature MFCC is calculated based on the energy statistics over 20 frequency bands~\cite{davis1980comparison}. Thus, it is more effective to recover the spectrum statistical features, instead of the spectrum details (i.e., each individual frequency component). Because the spectrum reduction attacks change the spectrum distribution (i.e., statistical features) by removing weak components~\cite{dft_attack}, they can finally reduce the inference accuracy. However, by adding magnitude-limited noise to the attacked audio, we can fix the statistical properties of the original spectrum. That is the root reason why the noise addition module can boost the inference accuracy, even though the added noise is different from the removed energy in the spectrum details.

% \vspace{0.03in}
{\noindent \bf Modeling.}
To add the Gaussian noise $n_{ns}(t)$ to the attacked audio $a(t)$, we directly process the signals in the time domain. Although frequency-domain processing is an alternative to achieve the noise addition module, time-domain signal processing is more efficient and does not need time-frequency conversion. Hence, we can get the recovered  signal $r(t)$ as
% \vspace{-0.05in}
\begin{equation}
    r(t) = a(t) + n_{ns}(t),
% \vspace{-0.05in}
\end{equation}
\noindent where $ns$ indicates the strength of the added noise, which is typically the standard deviation of the noise's magnitude distribution. We model the Gaussian noise by utilizing a random noise generator.

% \vspace{0.03in}
{\noindent \bf Adaptive Noise Addition.}
The selection of the noise parameter $ns$ can affect the performance of noise addition module. Strong noise (i.e., a large value of $ns$) can lead to the information loss of original audio signals. Meanwhile, noise that is too weak cannot compensate for the removed frequency components or restore the original statistical properties of the frequency spectrum. The optimal $ns$ value only depends on the component removal ratio of spectrum reduction attacks, which means we need to utilize different noise strengths for different component removal ratios.
For a specific component removal ratio, we test the inference accuracy of the attacked audio by applying noise addition module with different noise parameters. Thus, we can find the relationship between the component removal ratio and the best noise parameter $ns_{best}$. After the component removal ratio of an input speech is estimated, the noise addition module can adaptively select the best noise parameter to recover the original signal.

\subsection{Adaptation Module}

The \TN{} does not know the exact component removal ratio used by attackers in advance, but all predefined parameters in two core modules rely on component removal ratio to mitigate malicious audio optimally. To solve this issue, we propose an adaptive mitigation strategy by using the adaptation module.

The adaptation module can estimate the component removal ratio of input signals by calculating the proportion of extremely weak components. Specifically, in the \TN{} system, we calculate the ratio of the components, whose magnitude is less than 0.2\% of the max magnitude (acquired from experimental observations), among the whole spectrum.
% <= 0.0022886151071843342
Then, the adaptation module can further guide two core modules to set the optimal parameters according to the estimated component removal ratio. Therefore, with the adaptation module, we can achieve optimal mitigation performance adaptively. Note that the adaptation module can also distinguish if the input signals are malicious. If the estimated component removal ratio is near 0, the input signal is determined as benign and no additional operation is needed.

Attackers can only control the component removal ratio when performing spectrum reduction attacks; meanwhile, the adaptation module can estimate this parameter. Ideally, the~\TN{}~system can always obtain the optimal mitigation under a constant component removal ratio, even though attackers know the defense mechanism.
However, more armored attackers may irregularly change the attack parameter, which differs from the estimated one, to degrade the defense performance. For example, attackers can use random component removal ratios for different phonemes, words, or audio segments of a specific length. The minimum unit to change parameter depends on the minimal attack granularity (i.e., phoneme).

To defeat this adaptive attack method, the adaptation module is required to frequently detect the attack parameter and timely update the module parameters in \TN{}. Similarly, the minimum interval to update module parameters (e.g., $\alpha$) depends on the minimal segment length in our processing (i.e., $N$). Obviously, shorter segments can improve the system's adaptability and sensitivity, allowing it to adjust more frequently to the latest attack parameter set by attackers. However, we find that shorter segments also decrease the compensation performance (see Figure~\ref{fig:comp_NL}(a)) since the spectrum resolution decreases with a decrease of segment length. Thus, it is a trade-off to select a moderate segment length (i.e., update interval) for both performance and adaptability. However, this trade-off between adaptability and performance does not only restrict defenders but also put attackers in a dilemma. Finer attack granularity can help attackers switch the parameter more frequently, but it also reduces the error rate of attacked audio and impairs the attack effects (see Table~\ref{tab:adapt} in Section~\ref{sec:exp_adapt}).

% ------------
% \sw{add content: against adaptive attacks 
% [x]**adaptive attacker** who is aware of the defense. 
% [x]Even though the defense does not need to work perfectly under such conditions, it is interesting to see the limitations of the defense or if it can be easily circumvented.
% [x]Also, the attack is only evaluated for **one specific attack without variations**. 
% [x]The attacker can easily bypass the defense by adjusting some parameters, even randomly.}

% Second, in the compensation step, we evaluate the original component removal ratio.
% We will calculate the ratio of extremely weak components (i.e., ) among the whole spectrum.
\vspace{-0.05in}
\section{Evaluations}
\vspace{-0.05in}

\subsection{Experiment Setup}

% \vspace{-0.03in}
\noindent {\bf Runtime Environments.}
The \TN{} system is implemented using Python 3.10. All the evaluation experiments are conducted on a Linux (Ubuntu 20.04) server, equipped with an Intel Xeon 2620 CPU at 2.4 GHz and 16 GB RAM. The machine learning algorithms used in \TN{} are implemented by the \emph{scikit-learn 0.23} library only with the CPU resources. We achieve the functionality of speech recognition via the \emph{DeepSpeech 0.9.3}, \emph{PocketSphinx 5.0.0}, and \emph{webrtcvad 2.0.10} libraries. Also, the evaluation metric is implemented by the \emph{jiwer 2.3} library.

% \vspace{0.03in}
\noindent {\bf Speech Datasets.}
We select two benchmark datasets: TIMIT~\cite{TIMIT} and VCTK~\cite{VCTK}.
TIMIT Acoustic-Phonetic Continuous Speech Corpus is a standard dataset for automatic speech recognition~\cite{TIMIT}. The TIMIT corpus is collected from 630 female and male speakers with eight major American English dialects. Because each speaker read ten phonetically rich sentences, the dataset totally contains 6,300 audio samples, which consist of 4,620 training samples and 1,680 testing samples. No speaker is allowed to appear in both the training set and testing set. For each sample, the utterance is recorded as a 16-bit audio file with a sampling rate of 16 kHz. The dataset provides the ground-truth transcriptions of speech, as well as the time-aligned phonetic and word transcriptions. Due to the time alignment, it becomes convenient to achieve the phoneme-level and word-level spectrum reduction attacks in our experiments.

VCTK is a multi-accent corpus that includes the speech data of 110 English speakers~\cite{VCTK}. Each speaker reads about 400 sentences, which are stored at a sampling rate of 48 kHz. Since VCTK does not provide annotation for phonemes and words, we use 80 ms as average phoneme length and use 300 ms as average word length in the attack, based on our statistics on phonemes/words. In the real world, the fixed attack length is more practical since the attackers can save the enormous effort of locating phonemes/words. The phonetic/word segmentation can cost more time while does not have a unified standard~\cite{phoneticSegment}. We test the attack performance on TIMIT with the fixed segmentation length and find that the error differences are less than 4\% compared with the exact segmentation.

% \vspace{0.03in}
\noindent {\bf ASR Models.}
We select two open-source automatic speech recognition models DeepSpeech~\cite{deepspeech} and CMU Sphinx~\cite{CMUSphinx} in the inference stage.
The DeepSpeech model provides a speech-to-text engine using the deep learning techniques proposed by Baidu Research, supporting real-time inference on diverse platforms (e.g., Windows, macOS, Linux, Android, iOS) and embedded devices (e.g., Raspberry Pi). The acoustic model is trained on American English datasets, with the augmentation of noise synthesis.
DeepSpeech can achieve a word error rate of 7.06\% on the clean test set of LibriSpeech ASR corpus~\cite{Librispeech}.
The ASR model is imported via DeepSpeech library and runs only in a CPU environment, though DeepSpeech also supports GPU for a quicker inference. To detect voice activities in a piece of audio data, the webrtcvad library is used to find voiced segments and filter out unvoiced ones.
We also study the \TN{} performance via another ASR model CMU Sphinx.
Since the results are similar to those obtained via DeepSpeech, we list the detailed results in Appendix~\ref{append:perf}.
Note that the adversarial audios are generated for spectrum reduction attacks with different component removal ratios and the defense models are the same for attacks against two ASR models.

% \vspace{0.05in}
\noindent {\bf Spectrum Reduction Attack Settings.}
We reproduce the spectrum reduction attacks according to the workflow in Figure~\ref{fig:dft_attack}. For each audio sample in the datasets, we split the signal into several audio segments. Because the TIMIT dataset already marks the start and end sampling points of the phonetic and word transcriptions, we can precisely split the audio signals in phoneme/word-level granularity. We remove weak frequency components from each audio segment with a specific component removal ratio. For each audio segment, the DFT and IDFT operations are performed in the extended signal with a min length of the power of 2. Finally, we fill the processed audio segments into the original positions of the audio signal. According to the processing granularity, we reproduce both phoneme-level and word-level attacks. Also, as a more practical attack, we extend the spectrum reduction attack by setting a fixed segment length (80 ms or 300 ms). Compared with the original attack, audio is split into segments of equal length that are processed individually and filled into the original positions.

\begin{table*}[t]
\begin{center}
\caption{The performance of \TN{} and its each module against the word-level/phoneme-level spectrum reduction attacks (component removal ratio is 0.85). We evaluate both the WER and CER for the attacked audio and the audio with defense.}
% \vspace{-0.05in}
\label{tab:all}
\label{tab:comp}
\label{tab:noise}
\label{tab:echo}
\resizebox{.98\textwidth}{!}{
\begin{threeparttable}[b]
    \begin{tabular}{c|c|c|c|c|c|c|c}
    \toprule
    \multirow{2}{*}{\shortstack{\bf Dataset}} & \multirow{2}{*}{\shortstack{\bf Attack\\\bf Granularity}} & \multirow{2}{*}{\shortstack{\bf Evaluation\\\bf Metric$^{\dagger}$}} & \multirow{2}{*}{\shortstack{\bf Baseline\\\bf Error$^{\ddagger}$}} & \multirow{2}{*}{\shortstack{\bf Error w/\\\bf Attack$^{\S}$}} & \multicolumn{3}{c}{\bf Error w/ Our Defense$^{*}$} \\
    \cline{6-8}
    {} & {} & {} & {} & {} & {\bf Compensation} & {\bf Noise Addition} & {\bf \TN{}} \\
    \midrule
    % TIMIT
    \multirow{4}{*}{TIMIT} & \multirow{2}{*}{\shortstack{phoneme-\\level}} & {WER} & {0.217} & {0.597} & {0.336 (-68.7\%)} & {0.322 (-72.4\%)} & {0.314 (-74.5\%)} \\ 
    {} & {} & {CER} & {0.107} & {0.386} & {0.203 (-65.6\%)} & {0.190 (-70.3\%)} & {0.187 (-71.3\%)} \\ 
    \cmidrule{2-8}
    {} & \multirow{2}{*}{\shortstack{word-\\level}} & {WER} & {0.217} & {0.794} & {0.593 (-34.8\%)} & {0.570 (-38.8\%)} & {0.568 (-39.2\%)} \\ 
    {} & {} & {CER} & {0.107} & {0.562} & {0.396 (-36.5\%)} & {0.372 (-41.8\%)} & {0.370 (-42.2\%)} \\ 
    \midrule
    % VCTK
    \multirow{4}{*}{VCTK} & \multirow{2}{*}{\shortstack{phoneme-\\level}} & {WER} & {0.487} & {0.897} & {0.576 (-78.3\%)} & {0.641 (-62.4\%)} & {0.571 (-79.5\%)} \\
    {} & {} & {CER} & {0.375} & {0.705} & {0.419 (-86.7\%)} & {0.465 (-72.7\%)} & {0.415 (-87.9\%)} \\
    \cmidrule{2-8}
    {} & \multirow{2}{*}{\shortstack{word-\\level}} & {WER} & {0.487} & {0.885} & {0.691 (-48.7\%)} & {0.714 (-43.0\%)} & {0.686 (-50.0\%)} \\
    {} & {} & {CER} & {0.375} & {0.688} & {0.511 (-56.5\%)} & {0.522 (-53.0\%)} & {0.506 (-58.1\%)} \\
    
    \bottomrule
    \end{tabular}

    \begin{tablenotes}
        \item[$\dagger$] WER: word error rate between labels and predictions; CER: character error rate between labels and predictions.
        \item[$\ddagger$] Baseline Error indicates the average error rate when ASR infers original benign audio.  
        \item[$\S$] Error w/ Attack indicates the average error rate under spectrum reduction attack (including the baseline error).
        \item[$\star$] The percentage in parenthesis represents the reduction ratio to the errors caused by attacks.
    \end{tablenotes}

\end{threeparttable}
}
\end{center}
\vspace{-0.1in}
\end{table*}

\noindent {\bf Training Procedure.}
In the compensation module of \TN{}, the scaling coefficients need to be learned by linear regression. We use the close-form solution in Equation~\ref{eq:lr} to calculate the scaling coefficients over the shifted units. Considering all the samples in the dataset, the number of rows in matrix $H$ would be the total number of sampling points over all samples. However, the high dimensionality of $H$ will lead to the curse of dimensionality in the coefficient calculation, since Equation~\ref{eq:lr} involves the computation of $(H^T \cdot H)^{-1}$. Hence, to reduce the computing time and consumed resources, we divide all samples into multiple batches, where each batch contains 200 samples. We first calculate the scaling coefficients for each batch, i.e., $\alpha_{batch}$ and  then obtain the final coefficients $\alpha$ by calculating the mean vector of all the coefficients over batches, i.e., $\alpha = mean(\alpha_{batch})$. The selection of batch size is a trade-off between precision and speed. A larger batch size ensures the final coefficients are much closer to the real ones, while a smaller batch size can accelerate the model fitting speed.

% \vspace{0.05in}
\noindent {\bf Evaluation Metrics.}
We evaluate the performance of ASR inference transcriptions using word error rate (WER) and character error rate (CER), which are the common performance metrics for speech recognition and machine translation tasks~\cite{wer}. Given the ground truth and inference transcriptions, WER is the proportion of changed words (i.e., substitutions, deletions, and insertions) to the total words in the reference. WER measures the minimum edit distance between two transcriptions if we try to modify one towards another; thus, WER can also be used to indicate the similarity between two strings. Here, the minimum edit distance is also called Levenshtein distance, which is implemented by the Python module \emph{Levenshtein}. CER measures the similarity between two transcriptions in a finer character-level granularity since the attack is able to only change a few phonemes. WER cannot distinguish if the errors are caused by the whole words or part of phonemes; however, CER can reveal more phonetic information about the errors. For example, if the ground truth is `word' and we have two different transcriptions `wood' and `mode', the WERs will be the same since WER(`word', `wood') = WER(`word', `mode') = 1. However, CER can indicate the transcription `wood' can be better since CER(`word', `wood') = 0.25 while CER(`word', `mode') = 0.75.

To illustrate the error mitigation performance more directly, we define a new metric to indicate the change of WER (or CER) after the defense. For ASR systems, even benign audio has inherent word recognition error, which is denoted as $\text{WER}_0$. The word error rates of attacked audio and processed audio are denoted as $\text{WER}_a$ and $\text{WER}_p$, respectively. Therefore, $(\text{WER}_a - \text{WER}_0)$ is the error caused by attacks, while $(\text{WER}_a - \text{WER}_p)$ is the error mitigated by our defense. We define the \emph{WER reduction ratio} as $(\text{WER}_a - \text{WER}_p) / (\text{WER}_a - \text{WER}_0)$ to measure what proportion of errors caused by attacks are eliminated. Similarly, we define the \emph{CER reduction rate} to indicate how many character errors have been mitigated from the original attack. 
% The WER (CER) reduction rate can indicate how many errors have been mitigated from the original attack. 

% \vspace{0.03in}
% \noindent {\bf Ethics Consideration.}
% %
% We design the experiments only for the verification of the \TN{} performance, not for the malicious penetration of any specific ASR APIs.
% Besides the evaluation metrics, the researchers also analyze the inference errors by manually checking and summarization.
% However, the errors between the transcriptions are relatively objective.
% Therefore, our experiments do not involve any behavior analysis for humans.

% \subsection{Independent Module Analysis}

% performance of independent module and the parameter impact.

% \clearpage

%=========================================================
% \vspace{-0.03in}
\subsection{\TN{} Evaluation}

% \begin{table}[t]
% \begin{center}
% \caption{The performance of \TN{} against word-level and phoneme-level DFT attacks.}
% \label{tab:all}
% \renewcommand{\arraystretch}{1}
%     \begin{tabular}{c|c|c|c}
%     \toprule
%     {\bf DFT Attack} & {\bf Metric} & {\bf w/o \TN{}} & {\bf w/ \TN{}} \\
%     \midrule
%     \multirow{2}{*}{\shortstack{word-\\level}} & {WER} & {0.794} & {0.568 (-28.3\%)}  \\ 
%     {} & {CER} & {0.562} & {0.370 (-34.2\%)}  \\ 
%     \midrule
%     \multirow{2}{*}{\shortstack{phoneme-\\level}} & {WER} & {0.597} & {0.314 (-47.4\%)}  \\ 
%     {} & {CER} & {0.386} & {0.187 (-51.6\%)}  \\ 
%     \bottomrule
%     \end{tabular}
% \end{center}
% \end{table}

% \noindent {\bf Overall Performance.}
%
We conduct experiments to evaluate the effectiveness of \TN{} in defeating the spectrum reduction attacks. Since the \TN{} contains two modules (i.e., spectrum compensation and noise addition modules), we present the performance of each module in Sections~\ref{sec:comp} and~\ref{sec:noise} to illustrate the contribution of each module, respectively.

% \vspace{0.03in}
{\noindent \bf Performance of \TN{}.}
% 12
In Table~\ref{tab:all}, the phoneme-level spectrum reduction attacks on TIMIT cause a WER of 0.597 and a CER of 0.386 in the ASR inference results. However, with the mitigation of \TN{}, the WER and CER decrease to 0.314 and 0.187, respectively. The corresponding WER (CER) reduction ratio is 74.5\% (71.3\%) of the errors caused by attacks.
% 1-2
Note that CER is usually less than WER since CER only counts the characters that are inferred incorrectly, while WER counts the whole word even if only one character is incorrect.
% 34
For the word-level spectrum reduction attacks, the average WER and CER of attacked audio would be 0.794 and 0.562, respectively.
After utilizing \TN{}, the WER (CER) reduces to 0.568 (0.370) and its corresponding reduction ratio is 39.2\% (42.2\%).
% 12-34
Compared to the word-level attacks, the phoneme-level attacks cause a smaller error rate since taking phonemes as the basic units can better preserve the phoneme boundaries and hence lead to fewer shifted phoneme errors (refer to Section~\ref{sec:errortype}).
Also, the error reduction ratio against word-level attacks is less than that against phoneme-level attacks.
The reason is that word-level attacks typically cause a larger error rate due to the longer segment lengths, i.e., more points in DFT computation.

% use longer segment lengths (corresponding to more points in DFT computation), thus the frequency removal operations are in a finer granularity in the frequency spectrum.
% if use fixed segment length.
% When we perform the spectrum reduction attacks with a fixed segment length,

If being launched with a fixed segment length, phoneme-level attacks cause a WER (CER) of 0.581 (0.376), and word-level attacks cause a WER (CER) of 0.772 (0.541). With the \TN{}, the WER (CER) reduces to 0.304 (0.181) for phoneme-level attacks and reduces to 0.549 (0.358) for word-level attacks. The results show the error differences are typically less than 4\% when using a fixed phoneme/word length.

Next, we conduct the same experiments on another dataset VCTK. With the \TN{} system, the WER (CER) of phoneme-level attacks reduces from 0.897 (0.705) to 0.571 (0.415) while the WER (CER) of word-level attacks reduces from 0.885 (0.688) to 0.686 (0.506). The experiments show that the errors of VCTK samples are typically more than those of TIMIT samples, due to the different sources of the corpus. However, the error reduction ratio still ranges from 50.0\% to 87.9\%. Again, the experimental results prove that the \TN{} has mitigation effects against both the word-level and phoneme-level attacks on different benchmark datasets. We also evaluate the WER/CER for the compromised and mitigated audio using another ASR model, i.e., CMU Sphinx. The trends of mitigation effects are consistent with those tested by DeepSpeech. We show the results in Appendix~\ref{append:perf} and Table~\ref{tab:append_perf}.

Table~\ref{tab:all} also shows that these two modules have different mitigation effects on different data distributions, i.e., the spectrum compensation module performs better over the VCTK dataset and the noise addition module performs better over the TIMIT dataset. Thus, we apply two modules to achieve a better recovery effect.

% \vspace{0.03in}
{\noindent \bf Comparison with Baseline Error.}
Even without any attack, the original benign audio can lead to wrong transcriptions due to the fast speaking speed or rare words. Thus, we measure the WER and CER for the benign audio and set the error rates as the baseline error in our evaluation. The average WER (CER) of the benign audio is 0.217 (0.107) on TIMIT and 0.487 (0.375) on VCTK, which implies that recognition errors cannot be fully eliminated. Therefore, it is more meaningful to evaluate the error reduction capability of \TN{} over the errors caused by spectrum reduction attacks. On VCTK, the \TN{} can reduce the CER from 0.705 to 0.415 (drop by 0.29), while the max possible reduction is from 0.705 to 0.375 (drop by 0.33). Thus, only a 0.04 mitigable character error rate remains in the mitigated audio. In other words, we already eliminate 87.9\% errors among the mitigable errors. We analyze the error types for benign audio, attacked audio, and mitigated audio (in Section~\ref{sec:errorcomp}) and find the remaining errors in mitigated audio mainly come from the consonant errors, which are hard to be recovered due to the short duration and weak signal strength.

% We conduct the experiments by combining all these three modules together.
% The attacked audio will first pass through the compensation module and then be processed by noising module and echo module consecutively.
% In Table~\ref{tab:all}, we can find the performance is better than that of any individual module. 
% For word-level attack, the WER (CER) reduces to 0.568 (0.370) and the corresponding reduction ratio is (28.3\%) 34.2\%.
% For phoneme-level attack, the WER (CER) reduces to 0.314 (0.187) and the corresponding reduction ratio is 47.4\% (51.6\%).

% \vspace{0.03in}
{\noindent \bf Affecting Factors of \TN{}.}
The WER (CER) reduction ratio against the phoneme-level attacks is higher than that against the word-level attacks, due to two main reasons. First, for the phoneme-level attacks, the phonemes are processed independently. Thus, part of the recovered phonemes may lead to the correct ASR transcriptions since the NLP module can infer the words based on some critical phonemes. Second, the phoneme-level attacks preserve the phonetic boundaries, providing the possibility for \TN{} to have a better restoration effect. Besides, Table~\ref{tab:all} also shows the combination of modules can have better mitigation performance than any individual module. Therefore, it is necessary to include both the frequency domain compensation and time domain perturbations in the defense. Because \TN{} contains two core modules, we need to optimize a set of parameters for the module combination when encountering a different component removal ratio. There are 4 parameters in the \TN{} to be adjusted, thus we use the grid search~\cite{gridsearch} to get the optimized solution for different attack settings. By analyzing the reasonable range of parameters, we select a set of candidate values for each parameter. Given a component removal ratio, we obtain the optimal parameters by testing different parameter combinations. For example, when the component removal ratio is 0.85, we set the segment length $N$ to 32,768, the filter size $L$ to 45, the corresponding filter parameter vector $\alpha$, and the noise level $ns$ to 4.

% 32768
% 1278  971
% 0.0390  0.0296
% 1236  963
% 0.0367 0.02938
%=========================================================
% \vspace{-0.1in}
\subsection{Spectrum Compensation Module Evaluation}
\label{sec:comp}

We evaluate the performance of the spectrum compensation module and analyze the impacts of module parameters. 

% \vspace{0.03in}
\noindent{\bf Performance of Spectrum Compensation Module.}
Table~\ref{tab:comp} shows the phoneme-level spectrum reduction attack achieves a WER of 0.597 and a CER of 0.386 on the TIMIT inference results. However, with the mitigation of the spectrum compensation module, part of the removed frequency components can be recovered, and the WER and CER reduce to 0.336 and 0.203, respectively. The WER (CER) reduction ratio is 68.7\% (65.6\%) for phoneme-level attacks. For word-level attacks, the WER decreases from 0.794 to 0.593 and the CER decreases from 0.562 to 0.396. The WER (CER) reduction ratio is 34.8\% (36.5\%) for word-level attacks. To better understand the spectrum compensation, we compare the mean square errors (MSE) between the benign audio and the attacked/mitigated ones. The MSE reduces from 0.030 (0.039) to 0.027 (0.036) for phoneme-level (word-level) attacks. Indeed, the compensation also adds energy to strong components and may increase MSE (this part has less effect on the inference); however, the compensation for the removed components can greatly recover the missing energy hence reducing the MSE in total. On the VCTK dataset, we obtain similar results, where the error reduction ratio ranges from 48.7\% to 86.7\%. Thus, we conclude the compensation module is effective to mitigate both phoneme-level and word-level attacks. Moreover, the mitigation against phoneme-level attacks is much better since more than half of the inference errors can be fixed via spectrum compensation.

% \begin{table}[t]
% \begin{center}
% \caption{The performance of compensation module against word-level and phoneme-level DFT attacks.}
% \label{tab:comp}
% \renewcommand{\arraystretch}{1}
%     \begin{tabular}{c|c|c|c}
%     \toprule
%     {\bf DFT Attack} & {\bf Metric} & {\bf w/o Comp.} & {\bf w/ Comp.} \\
%     \midrule
%     \multirow{2}{*}{\shortstack{word-\\level}} & {WER} & {0.794} & {0.593 (-25.3\%)}  \\ 
%     {} & {CER} & {0.562} & {0.396 (-29.6\%)}  \\ 
%     \midrule
%     \multirow{2}{*}{\shortstack{phoneme-\\level}} & {WER} & {0.597} & {0.336 (-43.7\%)}  \\ 
%     {} & {CER} & {0.386} & {0.203 (-47.5\%)}  \\ 
%     \bottomrule
%     \end{tabular}
% \end{center}
% \end{table}

% Loading file: .//logs/log_sig_R0.85_N32768_L45.txt
% WER1: ATK 0.7935401511938656, CMP 0.5926393058647975, RATE 25.317035946677258%
% CER1: ATK 0.5618050322665629, CMP 0.39571587098351085, RATE 29.563487641429088%
% WER2: ATK 0.5970629969807137, CMP 0.336407670333089, RATE 43.65625201456663%
% CER2: ATK 0.3863131146995542, CMP 0.20288972819087514, RATE 47.48049691539264%

\begin{figure}[t]
    \centering
    \subfloat[phoneme-level attack]{
    \includegraphics[width=0.495\linewidth]{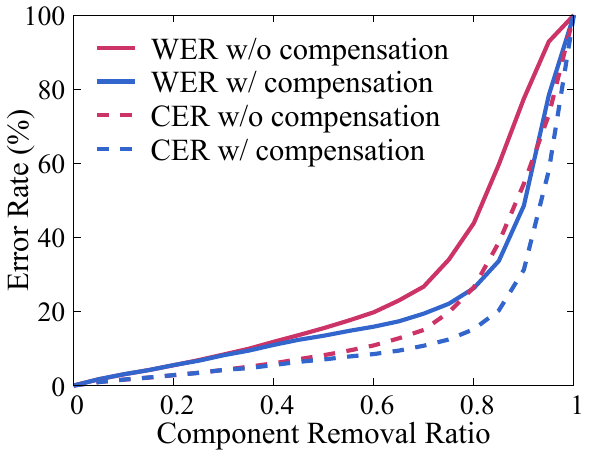}
    }% 
    \subfloat[word-level attack]{
    \includegraphics[width=0.495\linewidth]{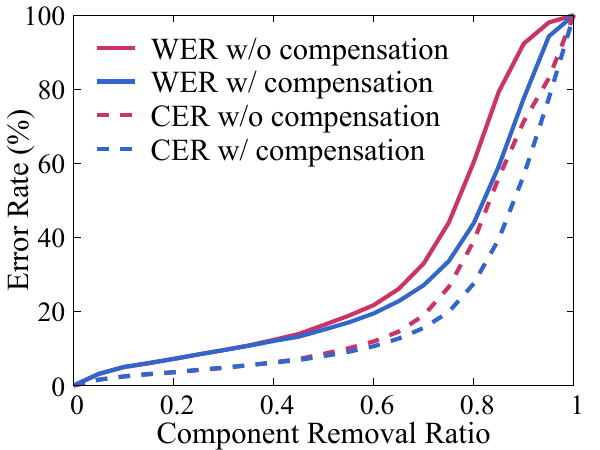}
    }% 
    % \vspace{-0.05in}
    \caption{Performance of spectrum compensation module against phoneme-level and word-level attacks with different component removal ratios.}
    \label{fig:comp_R}
    \vspace{-0.2in}
\end{figure}

% \vspace{0.03in}
\noindent{\bf Impact of Component Removal Ratio.}
We test the spectrum compensation module against spectrum reduction attacks with different component removal ratios. Figure~\ref{fig:comp_R} shows that the inference error rate changes under different attacks. We find similar trends of WERs and CERs over the component removal ratios for both phoneme-level and word-level attacks. With the increase of component removal ratio, both WER and CER increase from 0\% to 100\%. This trend is intuitive since more information is deleted with a higher component removal ratio. Compared with the word-level attacks, the phoneme-level attacks cause lower WER and CER. From the perspective of the frequency domain, a phoneme-level signal is shorter and the corresponding DFT spectrum has fewer units; thus the component removal operations become rougher. Besides, ASR can still have a chance to infer the phonemes correctly by phonetic context information. Also, we find the WER (CER) reduction ratio is higher when the component removal ratio is between 0.5 to 0.9. Too low component removal ratio leads to few total errors; while too high component removal ratio hinders the compensation due to the little residual information. However, the spectrum compensation model can always reduce the error rate even if the component removal ratio may vary.

\begin{figure}[t]
    \centering
    \subfloat[error reduction rate vs. N]{
    \includegraphics[width=0.495\linewidth]{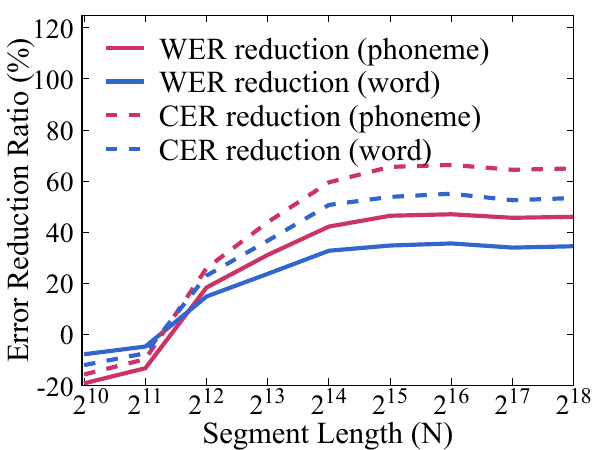}
    }% 
    \subfloat[error reduction rate vs. L]{
    \includegraphics[width=0.495\linewidth]{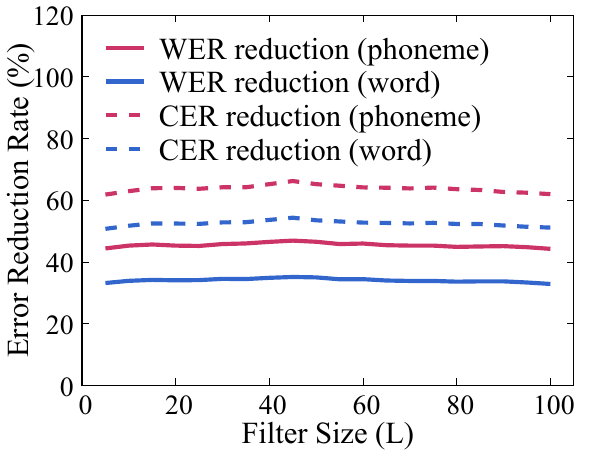}
    }% 
    % \vspace{-0.05in}
    \caption{Performance of spectrum compensation module against phoneme/word-level attacks over different segment lengths and filter sizes (component removal ratio is 0.85).}
    \label{fig:comp_NL}
    \vspace{-0.2in}
\end{figure}

% \vspace{0.03in}
\noindent{\bf Impact of Segment Length.}
To evaluate the impact of module parameters, we first conduct the controlled-variable experiments on the segment length ($N$) of audio signals. To facilitate the DFT computation, the length of signal segment is usually in the form of a power of two. Thus, we test the performance of spectrum compensation module with a segment length range from $2^{10}$ to $2^{18}$ sampling points. Considering the sampling rate ($f_s$) of audio signals is 16 kHz, the audio segment ranges from 64 ms to 16 sec. Given a fixed component removal ratio of 0.85, Figure~\ref{fig:comp_NL}(a) shows the error reduction ratio on both phoneme-level and word-level attacks with different segment lengths. We can find the error reduction ratio increases as the $N$ increases, which means a larger $N$ can provide a better compensation performance. The performance becomes stable when $N \geq 2^{15}$. Because too large $N$ can also increase the computation complexity of DFT, we select $2^{15}$ (i.e., 2s) as the optimal parameter for the compensation module. Note that when $N$ is too small, the compensation module may also hurt the inference results. If $N \leq 2^{11}$, the signals need to be concatenated frequently and the sampling points near segment boundaries cannot be well learned. Also, short signal segments may split too many long phonemes, hence increasing the phonetic ASR inference errors. Hence, $2^{15}$ is a trade-off between computation complexity and module performance.

% \vspace{0.03in}
\noindent{\bf Impact of Filter Size.}
We also conduct controlled-variable experiments on the filter size ($L$) of the spectrum compensation module. We set the filter size range from 5 to 100 with a step of 5 and use the compensation module to mitigate the spectrum reduction attacks. Figure~\ref{fig:comp_NL}(b) shows the WER (CER) reduction ratio using different filter sizes, under both the phoneme-level and word-level attacks with a fixed component removal ratio of 0.85. From the experimental results, we can find the critical information to infer the removed components are concentrated on the neighboring components. Therefore, the spectrum compensation module still performs well even with a filter size of 5. When $L \leq 45$, the mitigation performance increases moderately as the $L$ increases since more useful neighboring information is involved. However, when $L \geq 45$, the mitigation performance decreases slightly as the $L$ increases since the involved irrelevant noise has negatively affected the model fitting. Also, too large $L$ value will increase the dimensionality of $H$ in Equation~\ref{eq:lr}, hence increasing memory consumption. Thus, based on all factors, we set the optimal filter size as 45.

% \clearpage
%=========================================================
% \vspace{-0.05in}
\subsection{Noise Addition Module Evaluation}
\label{sec:noise}

% \vspace{-0.02in}
\noindent{\bf Performance of Noise Addition Module.}
Table~\ref{tab:noise} shows the noise addition module can reduce the WER (CER) to 0.322 (0.190) against the phoneme-level attacks on TIMIT. The WER (CER) reduction ratio is 72.4\% (70.3\%) for the phoneme-level attacks. For the word-level attacks, the WER (CER) decreases to 0.570 (0.372) and the corresponding reduction ratio is 38.8\% (41.8\%). On the VCTK dataset, the experiments show similar results, where the error reduction ratio ranges from 43.0\% to 72.7\%. We find the noise addition module is highly efficient to defeat both the phoneme-level and word-level attacks and half of the introduced inference errors are alleviated via adding adaptive noise. The experimental results also reflect that the removed components during the spectrum reduction attacks have a similar statistical property to the Gaussian noise.

% \begin{table}[t]
% \begin{center}
% \caption{The performance of noising module against word-level and phoneme-level DFT attacks.}
% \label{tab:noise}
% \renewcommand{\arraystretch}{1}
%     \begin{tabular}{c|c|c|c}
%     \toprule
%     {\bf DFT Attack} & {\bf Metric} & {\bf w/o Noise} & {\bf w/ Noise} \\
%     \midrule
%     \multirow{2}{*}{\shortstack{word-\\level}} & {WER} & {0.794} & {0.570 (-28.2\%)}  \\ 
%     {} & {CER} & {0.562} & {0.372 (-33.8\%)}  \\ 
%     \midrule
%     \multirow{2}{*}{\shortstack{phoneme-\\level}} & {WER} & {0.597} & {0.322 (-46.0\%)}  \\ 
%     {} & {CER} & {0.386} & {0.190 (-50.9\%)}  \\ 
%     \bottomrule
%     \end{tabular}
% \end{center}
% \end{table}

% Load data from: .//logs/log_ns_R0.85_NS6.txt
% -------------------------------------
% WER1: ATK 0.7935401511938656, CMP 0.5697920509825279, RATE 28.196191443459163%
% CER1: ATK 0.5618050322665629, CMP 0.37218325094289917, RATE 33.75223973317718%
% WER2: ATK 0.5970629969807137, CMP 0.3223855750018195, RATE 46.0047638805134%
% CER2: ATK 0.3863131146995542, CMP 0.18970460781782109, RATE 50.8935626052045%

\begin{figure}[t]
    \centering
    \subfloat[error rate vs. attack parameter]{
    \includegraphics[width=0.495\linewidth]{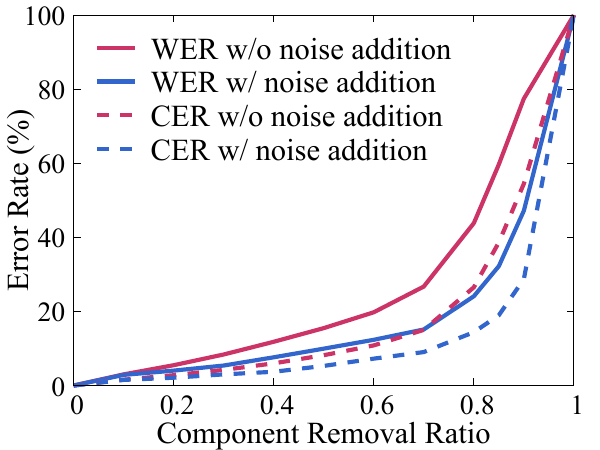}
    }% 
    \subfloat[error reduction rate vs. $ns$]{
    \includegraphics[width=0.495\linewidth]{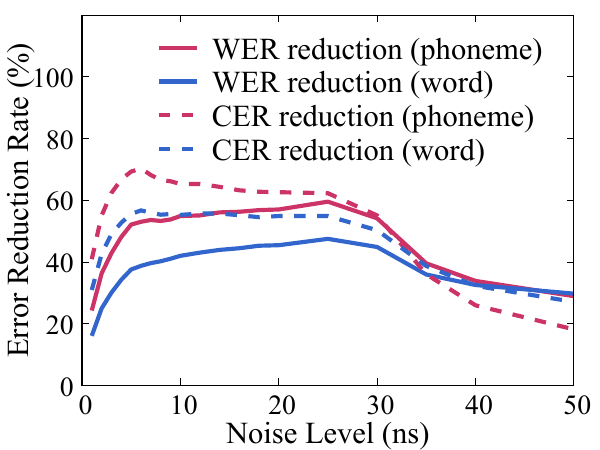}
    }% 
    % \vspace{-0.05in}
    \caption{The performance of noise addition module against the spectrum reduction attacks with different component removal ratios and noise levels.}
    \label{fig:noise_paras}
    \vspace{-0.2in}
\end{figure}

% \vspace{0.03in}
\noindent{\bf Impact of Component Removal Ratio.}
We test the noise addition module against the attacks with different component removal ratios. Figure~\ref{fig:noise_paras}(a) shows the WERs (CERs) of the attacked audio with and without the noise addition module, under the phoneme-level attack. The noise level in the module (i.e., the noise standard deviation) is adapted to the component removal ratio. Our results show a larger component removal ratio requires a larger noise parameter since more energy is needed to recover the spectrum distribution. If we use a fixed noise level, the added noise can recover part of the removed components and reduce both the WER and CER of attacked audio, only if the component removal ratio is greater than a specific value. However, if the component removal ratio is less than this threshold, the added noise can negatively affect the audio and even increase the WER and CER since the added energy is greater than the removed one. Thus, the adaptation module adopts a smaller noise level for a smaller component removal ratio. We also find Gaussian noise is an appropriate perturbation to reduce the success rate of phoneme-level attacks. The results also hold for word-level attacks.

% Moreover, we find the adaptive noise level will increase as the component removal ratio increases in the experiments.

% When the component removal ratio is greater than a specific value (0.5 in this case), the added noise can recover part of the removed frequency components and reduce both the WER and CER for attacked audio.

% For a larger component removal ratio, the noise level must be correspondingly larger.

% In Figure~\ref{fig:noise_paras}(a), we set the noise level (i.e., the standard deviation of noise) as a fixed value of 6 and obtain the WERs (CERs) under the attacks with different component removal ratios.
% Different from the compensation module, a fixed noise level does not always bring mitigation to the attacked audio.

% \vspace{0.05in}
\noindent{\bf Impact of Noise Level.}
To analyze the impact of different noise levels, we conduct the experiments by performing the attacks with a fixed component removal ratio of 0.85. In Figure~\ref{fig:noise_paras}(b), we can find the WER (CER) reduction ratio first increases and then decreases, with the increase of noise level. Weak noise is not enough to compensate for the defect in the spectrum distribution, hence having a limited effect on error reduction. However, too much noise will introduce excessive signal interference and overwhelm the original spectrum distribution; thus the effect of error reduction will be gradually eroded with a higher noise level. Therefore, for a specific component removal ratio, we need to find the optimal noise level, which can lead to the best error reduction effect.

% \vspace{-0.03in}
\subsection{Adaptation Module Evaluation}
\label{sec:exp_adapt}
% 80ms ~ 5000ms
% 512(80)-160-320-640-1280-2000-5000
To evaluate the effectiveness of the adaptation module, we test the error mitigation performance of~\TN{} under a dynamic attack environment, where attackers periodically change the attack parameter. We find the~\TN{} performance is stable when $N=2^{15}$, under different spectrum removal ratios. Therefore, we use a fixed $N$ value of $2^{15}$ ($\sim$2s), which is the min update interval of module parameters. For the attack settings, we set the min attack granularity as 80 ms, i.e., the average phoneme duration. The max attack granularity is set to $2N$ ($\sim$4s). Our update interval ($N$) can handle the attack granularity larger than $2N$ since our update frequency will be more than twice of attackers'. When launching the attacks, attackers randomly change the spectrum removal ratio as a uniform distribution, i.e., $R\sim U(0.01,0.95)$, according to the attack granularity. When $R=0.01$, only 1\% components are removed; when $R>0.95$, the attacks will affect human comprehension.

We use TIMIT as the test dataset, which has an inherent CER of 10.7\%. Table~\ref{tab:adapt} shows the CERs of attacked audio and mitigated audio, under different attack unit lengths (granularity). With the increase of attack unit length, the CER of attacked audio will increase until converging into a certain value, which is consistent with the fact that word-level attacks perform better than phoneme-level attacks. Therefore, we find attackers also suffer from a trade-off, where a smaller attack unit can increase the parameter changing frequency while decreasing the attack performance. With a smaller attack unit, the CER of mitigated audio tends towards the minimum limit of CER (i.e., 10.7\%), causing a relatively higher error reduction ratio. Also, the CER of mitigated audio converges when the attack unit becomes larger, leading to a convergent error reduction ratio. The experimental results show the adaptation module can still perform well even if attackers use arbitrary attack units and random component removal ratios.

% ----------
% \sw{Notes:
% N selection
% to evaluate the adaptation performance -> different attack granularity (min unit to change attack parameter).
% average phoneme 80ms
% while N=(2s)
% from granularity from 80ms to twice of N (4s).
% when larger than 2N, cover.
% ----
% table we can find...}

\begin{table}[t]
\begin{center}
\caption{The performance of~\TN{}~system under a dynamic attack environment with different attack granularities.}
\label{tab:adapt}
% \vspace{-0.1in}
\renewcommand{\arraystretch}{1}
\resizebox{0.99\linewidth}{!}{
    \begin{tabular}{c|c|c|c|c|c|c|c}
    \toprule
    attack unit (ms) & 80 & 160 & 320 & 640 & 1280 & 2000 & 4000 \\
    \midrule
    {CER w/ attack (\%)} & {16.9} & {19.1} & {18.3} & {23.8} & {22.1} & {24.0} & {23.2} \\ 
    {CER w/ \TN{} (\%)} & {11.8} & {13.7} & {14.1} & {19.3} & {17.4} & {19.1} & {18.4}  \\ 
    \midrule
    {error reduction (\%)} & {82.3} & {64.3} & {55.3} & {34.4} & {41.2} & {36.8} & {38.4}  \\ 
    \bottomrule
    \end{tabular}
}
\end{center}
\vspace{-0.1in}
\end{table}

\subsection{{Overhead}}

Since content moderation systems are often deployed at scale to process a significant amount of content traffic, the performance cost of \TN{} should be low for practical usage.
The \TN{} system only takes 40 ms to process 1-second audio online.
% The processing time scales proportional to the audio signal length, so finer signal segmentation can reduce the processing time.
Meanwhile, since the audio is processed as a data stream, this processing delay is imperceptible to human listeners.  
Moreover, the peak RAM usage of \TN{} is 230 MB, measured by the \emph{memory\_profiler} package~\cite{memprofiler}.
Since the basic operations in our algorithms are discrete Fourier transform, matrix multiplication, and vector addition, the system can be further accelerated and optimized via dedicated hardware, parallel computing, and optimized C code.

\section{Residual Error Analysis}
% \vspace{-0.06in}
Though \TN{} can successfully remove the majority of errors introduced by the spectrum reduction attacks, some ASR inference errors still remain in the transcriptions. To better understand the remaining errors, we launch an error analysis. We manually check the inference errors for 1,680 test samples, each of which has the forms of benign, attacked, and mitigated audio. We reproduce both the phoneme-level and word-level spectrum reduction attacks with a specific component removal ratio of 0.85. In total, we check 1,680 benign audio, 3,360 attacked audio, and 3,360 mitigated audio. We summarize six types of errors from the ASR inference results and analyze the cause of errors. 
In fact, these errors exist in general ASR applications and are not limited to specific audio attacks. % including the spectrum reduction attacks.

%mark the error types for each audio transcription. Finally, we analyze the error composition of different audio transcriptions and provide insight into future defense design\sw{s against voice attacks}.

% \vspace{-0.05in}
\subsection{Types of ASR Inference Errors}
% \vspace{-0.01in}

\label{sec:errortype}

{\noindent \bf T1: Fast Speed (Elision) Errors.} 
A common inference error type is caused by fast speaking speed. Human speakers may unconsciously omit one or more phonemes in a word or phrase~\cite{elision}. However, because most of the ASR training samples are clear audio, the test audio with elision might not fall into the regular inference distribution. Hence, even the original audio without any attack might not be recognized by ASR, e.g., the following transcription (T) misses the phrase `don't ask me' compared with the ground truth (G).
% \vspace{-0.06in}
\begin{Verbatim}[commandchars=\\\{\},fontfamily=courier,fontsize=\small]
G: \CASE{don't ask me} to carry an oily rag like that.
T: to carry an oily rag like that.
\end{Verbatim}
% \vspace{-0.06in}
These errors usually occur in the sentence beginning due to no prompt ahead for inference. Humans have the everyday experience of using phrase elision and hence can recognize fast-speaking sentences. The elision errors cannot be solved by simply adding time-compressed audio to the training set since the time compression can also increase the phoneme frequencies; while the fast-speaking audio only shortens the duration of phonemes but does not change their fundamental frequencies. However, this error type can be further mitigated by adding real spoken corpus into ASR training set~\cite{FakeWake}.

% \vspace{0.03in}
{\noindent \bf T2: Rare Word Errors.} 
If the involved words are rarely used in daily life, there will be a high probability of recognition error. In the ASR, rare words have lower prior probabilities, hence resulting in low  weights in the inference outputs. For example, `iguanas' and `alligators' are two obscure species names, which lead to errors in the transcription.
% \vspace{-0.25in}
\begin{Verbatim}[commandchars=\\\{\},fontfamily=courier,fontsize=\small]
G: \CASE{iguanas and alligators} are tropical reptiles.
T: \CASE{quanta analogous} are tropical reptiles.
\end{Verbatim}
% \vspace{-0.05in}
This error type is due to the intrinsic inference mechanism of ASR systems and can be mitigated by meta-learning~\cite{meta_rare} or backoff strategy~\cite{backoff_rare}.

% \vspace{0.03in}
{\noindent \bf T3: Consonant Errors.}
The consonant confusion is usually caused by different accents. It is easy to confuse [b] and [p], [t] and [d], or [m] and [n] due to the short duration and similar pronunciation. However, consonant errors are also common errors in ASR systems. The following example shows the misinterpreted consonants in transcription, especially for the attacked audio. Also, ASR may not catch specific consonant phonemes with unstressed pronunciation, e.g., the stop sounds [t], [d], and [k] at the end of words. 
% \vspace{-0.24in}
\begin{Verbatim}[commandchars=\\\{\},fontfamily=courier,fontsize=\small]
G: the one \CASE{m}ea\CASE{t} showing .. a\CASE{t} .. dose\CASE{s} is por\CASE{k}.
T: the one \CASE{n}ee\CASE{d} showing .. an\CASE{d} .. does is poor.
\end{Verbatim}
% \vspace{-0.05in}

% \vspace{0.03in}
{\noindent \bf T4: Vowel Errors.}
There are 15 vowels in American English. Different from the consonants, vowels are made with the mouth fairly open; thus vowels are louder and do not need precise articulation like consonants. Hence, vowels are much less likely to be misinterpreted. However, with the spectrum reduction attacks, vowels can still be incorrectly recognized since the spectrum has been changed. Vowels may confuse each other due to different vowel blackness (e.g., the front vowel [e] and the central vowel [\textschwa]) or vowel heights (e.g., the close-mid vowel [o] and the close vowel [u]).
% \vspace{-0.08in}
\begin{Verbatim}[commandchars=\\\{\},fontfamily=courier,fontsize=\small]
G: w\CASE{i}ll robin wear a .. sh\CASE{owe}d pleasure.
T: w\CASE{e}ll robin where a .. sh\CASE{oul}d pleasure.
\end{Verbatim}
% \vspace{-0.06in}

% \vspace{0.03in}
{\noindent \bf T5: Shifted Phoneme Errors.}
The shifted phoneme error means the phonemes in a word are broken up into two or more words, e.g., the last few phonemes in a word are tied to the next word.
This error type is usually a derivative of consonant and vowel errors.
In this example, ASR does not catch [f] in the word `fairy', thus the [\texttheta] in `tooth' has been shifted to `-airy' to form a new word `theory'.
% \vspace{-0.06in}
\begin{Verbatim}[commandchars=\\\{\},fontfamily=courier,fontsize=\small]
G: the \CASET{too}\CASE{th fairy} \CASET{for}\CASE{got} to .. \CASET{too}\CASE{th fell} out.
T: the \CASET{two} \CASE{theories} \CASET{for} \CASE{that} to .. \CASET{to} \CASE{sell} out.
\end{Verbatim}
% \vspace{-0.05in}
Sometimes, the shifted phonemes can be simple, e.g., `summertime' is translated as `summer time'.
In this case, the error only affects the WER but has no changes to the CER.

% \vspace{0.03in}
{\noindent \bf T6: NLP Inference Errors.}
The natural language processing (NLP) module in the ASR can infer words according to the context information. 
Hence, with a predefined language model, the NLP module can output an inference result that might not be the same as the real sentence.
In the following example, the NLP module tends to keep the personal pronoun identical and changes the word `your' to `her'.
This error is mainly caused by the insufficient context information of the test sample.
% \vspace{-0.06in}
\begin{Verbatim}[commandchars=\\\{\},fontfamily=courier,fontsize=\small]
G: she had \CASE{your} dark suit \CASE{in} greasy wash water.
T: she had \CASE{her} dark suit \CASE{and} greasy wash water.
\end{Verbatim}
% \vspace{-0.06in}
The NLP inference errors are sometimes induced by other inference errors. 
Other errors may change the critical words in the sentences, hence changing the NLP inference results.

% \vspace{-0.075in}
\subsection{Error Composition Analysis}
% \vspace{-0.03in}

\label{sec:errorcomp}

\begin{figure}
    \centering
    \includegraphics[width=0.95\linewidth]{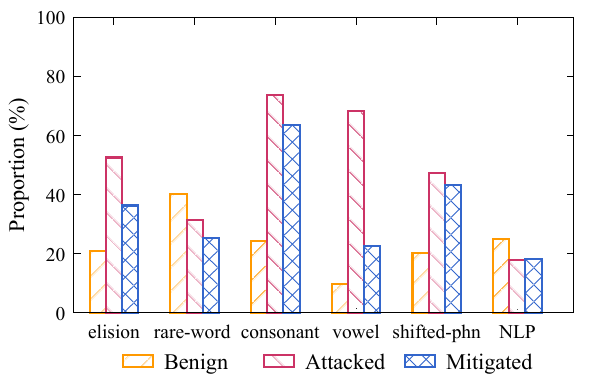}
    % \vspace{-0.1in}
    \caption{The proportion of six types of inference errors in wrong samples for the benign, attacked, and mitigated audio.}
    \label{fig:case}
    \vspace{-0.2in}
\end{figure}

Based on the above six error types, we check the inference errors in the test data. In Figure~\ref{fig:case}, we illustrate the proportion of each error type over the samples of wrong transcriptions. Because one transcription may have multiple error types, the sum of all the proportion values can be larger than 100\%.

% \vspace{0.03in}
{\noindent \bf Errors of Benign Audio.}
Indeed, even a benign audio without any modification has some ASR inference errors. By utilizing the DeepSpeech ASR API, we test the WER/CER for the audio in the original dataset. The experimental results show that the WER (CER) of the benign audio is 0.217 (0.107) on TIMIT, which means 20\% of the inferred words need to be corrected. In Figure~\ref{fig:case}, we analyze the error types for the benign audio. We find the errors mainly come from the rare word errors (40\%) and NLP inference errors (25\%). However, these errors are inherent due to the ASR training corpus and the NLP language model; hence, the errors cannot be further eliminated by post-processing. Moreover, these errors are general for ASR systems and are not specific to any attacks. Due to the clear pronunciation, the benign audio has a much lower error rate of elision errors, compared to the attacked audio. Also, for the benign audio, the phoneme errors (i.e., consonant errors, vowel errors, and shifted phoneme errors) are relatively fewer.

% \vspace{0.03in}
{\noindent \bf Errors of Attacked Audio.}
The WER (CER) increases to 0.597 (0.386) for phoneme-level attacks and 0.794 (0.562) for word-level attacks. In Figure~\ref{fig:case}, we find the error distribution has also been largely changed. The two main error sources are consonant errors (74\%) and vowel errors (68\%), then the elision errors (53\%) and shifted phoneme errors (47\%). Obviously, these phoneme errors have dominated the inference results. The root cause is that spectrum reduction attacks have changed the spectrum distributions of phonemes by removing weak frequency components. Hence, based on the statistical features (e.g. MFCC~\cite{davis1980comparison}), the ASR is apt to misinterpret the attacked audio in the minimum sound units, i.e., phonemes. Besides, the spectrum reduction attacks blur the boundaries between phonemes; thus, there are more elision errors in the transcriptions. The proportion of rare word errors decreases, not because the attack can improve the rare word recognition, but because ASR does not even provide the transcriptions for these words thus we do not count for this error type. Above all, adversarial audio mainly works by causing phoneme errors.

% \vspace{0.03in}
{\noindent \bf Errors of Mitigated Audio.}
After applying the \TN{} system, we find the proportions of most error types have decreased. However, compared with the benign audio, there is still a gap in the error rate. The WER (CER) of mitigated audio is 0.314 (0.187) for phoneme-level attacks and 0.568 (0.370) for word-level attacks. Considering part errors come from the inherent errors in the benign audio, the actual surplus of WER (CER) is 0.097 (0.080) for phoneme-level attacks and 0.351 (0.263) for word-level attacks. Within the feasible extent of error mitigation, the WER (CER) actually reduces by 74.5\% (71.3\%) for phoneme-level attacks and 39.2\% (42.2\%) for word-level attacks. Figure~\ref{fig:case} shows most of the reduced errors are in the vowel and elision error types. Vowel phonemes have higher loudness and signal strength; thus, the attacked vowels can be recovered more easily based on more information. However, many consonant phonemes are light sounds and have a shorter duration, so audio attacks are more likely to affect the critical features that are hard to recover. That is why consonant errors and shifted phoneme errors are still the major errors of mitigated audio. \TN{} alleviates the elision errors by recovering most of the vowel phonemes, and future mitigation should focus on the recovery of consonant phonemes.
% \vspace{-0.05in}
\section{Discussions}

\subsection{Multipath Effect and Audio Quality Improvement}
Multipath effect is a practical perturbation when the audio propagates in an over-the-air environment. The arrival signal is a combination of the signals coming from different paths by reflection and scattering. By designing an echo module, we analyze if multipath perturbations can mitigate spectrum reduction attacks. The details of the echo module are listed in Appendix~\ref{append:multipath}. However, the experimental results show the echo module only provides a minor improvement in error reduction. Although both noise and multipath effects are acoustic propagation effects, the latter does not introduce components of new frequencies into the attacked audio to fill the removed components. Hence, we exclude the echo module for \TN{}. 

In addition, the audio quality can be improved by denoising (e.g., Wiener filter~\cite{chen2006new}), echo cancellation (i.e., inverse of our echo module), or equalizer (i.e., volume adjustment for each frequency~\cite{bruschi2021low}); however, since these methods cannot introduce new frequency components into the compromised audio, they are limited on mitigating spectrum reduction attacks.

%\vspace{-0.05in}
\subsection{Usability}
\TN{} can be deployed on the devices and services containing automatic speech recognition function to perform content moderation. When attackers use spectrum reduction attacks to spread harmful information to public  and escape the content moderation systems, \TN{} can recover most of the affected words/characters. \TN{} aims to reduce WER/CER to the baseline level, even it is hard due to the irreversible information loss; however, reducing the flipped words/characters can still benefit content moderation systems by inferring harmful content from a more accurate context. Moreover, \TN{} can be extended to mitigate other audio attacks. For the audio attacks that directly change signal spectrum, our spectrum compensation module can be applied to recover the original spectrum, while the only change would be the new coefficients trained on new attacked audio. The noise addition module and the echo module are used to model the acoustic propagation process, which is effective against multiple adversarial attacks. %Compared with the physical-based defense (i.e., over-the-air replay and record), the \TN{} system can process the audio signals faster while achieving satisfying mitigation effects.

\TN{} can be either embedded into hardware (e.g., digital signal processor) or implemented by software (e.g., the signal preprocessing module in the ASR pipeline). Since the adaptation module can  estimate the attack parameters dynamically, \TN{} does not affect the ASR function under no attack. Besides the detection and prevention functions that can be provided by a classifier, \TN{} can further provide recovery function, which is critical to content moderation systems that need original contents for further processing, e.g., timely reporting terrorism.

% Also, some attackers may use spectrum reduction attacks to hide their voice communications in a controlled network so that their sensitive keywords cannot be noticed by the detection system.
%

% **1-Can we improve audio quality to defeat spectrum reduction attacks?**
% Audio quality can be improved by denoising (band-pass filter), echo cancellation (inverse of our Echo module), or equalizer (adjusting volume of each frequency); however, these methods cannot introduce new frequency components into the compromised audio, hence having limited effects on mitigating spectrum reduction attacks. We will clarify it in Section-7.1.

\subsection{Countering Attack Variants}

In Figure~\ref{fig:dft_attack}, spectrum reduction attacks are shown in the basic form; however, attackers may launch attack variants by changing either the attack frequency scope or the remaining component magnitude.
% attack variant-1.
First, attackers may launch spectrum reduction attacks only over specific frequency bands, instead of the entire spectrum. The attack on the local spectrum leads to a mismatch of the evaluated $R$ value; however, its effect on our defense is limited since the mismatch effect is similar to the adaptive attacks that frequently change $R$ values, while our system is insensitive to the mismatch of $R$ value estimation (see Table~\ref{tab:adapt}). Besides, we can detect and only compensate for the affected frequency bands. This attack may also impair attack effects since the changed local spectrum may not alter the total distribution and affect the final inference results.

Second, attackers may try to attenuate the energy to a low level (e.g., 50\%) instead of to zeros. This attack leads to a smaller estimated $R$ value in our defense hence reducing the compensated energy. However, we do not need to compensate for all energy loss since there is still part of the energy retained in the components. In Appendix~\ref{append:half_energy}, we conduct the experiments with the new attack settings that retain 25\%, 50\%, and 75\% energy, respectively.
The experiments show the CER reduction rate is 57.0\% on VCTK for word-level attacks with 50\% retained energy, compared to 58.1\% in normal settings.
Therefore, our defense is robust against this attack variant.

% What if attackers perform spectrum reduction on specific frequency bands?
% First, the attack on the local spectrum can lead to a mismatch of R-values over different frequency bands; however, its effect on our defense is limited since this mismatch is similar to the adaptive attacks that frequently change R-values. Meanwhile, our system is insensitive to the mismatch of R-value estimation (see Table-2). 
% Second, Section-5.5 shows attackers will impair their attack effects when trying to bypass our defense. When attackers focus on specific frequency bands, the local spectrum changes may not alter the total distribution and affect the inference results. 
% Besides, we can detect the affected frequency bands and only compensate for these frequency bands.

% What if attackers reduce the energy of frequency points to 50\% instead of to zeros? 
% First, this attack method will impair the attack performance since the statistical distribution may not be sufficiently changed.
% Second, it can hardly affect our defense. When reducing the energy of some frequency points to 50\%, the attack leads to a smaller estimated component removal ratio R of our defense that reduces the compensated energy. Thus, we don’t need to compensate for all energy loss since there is 50\% of energy retained. We conduct new experiments that show the CER reduction rate is 57.0\% on VCTK for word-level attacks, compared to 58.1\% in normal settings. We will clarify it in Section-7. 

% attack var2

\vspace{-0.05in}
\subsection{Limitations}
% \vspace{-0.05in}
%

Our system cannot fully eliminate recognition errors mainly due to two reasons. First, some inherent errors caused by ASR and corpus selection even exist in benign audio and cannot be fully eliminated by post-processing. Second, spectrum reduction attacks decrease the amount of information by removing weak components, so the irreversible increase of information entropy  hinders the recovery of original signals.

The \TN{} performance relies on parameter selection, computing resources, and segment granularity. First, \TN{} contains 4 parameters (i.e., 3 in spectrum compensation module and 1 in noise addition module), making it time-consuming to find the optimal combination. However, since all parameters are related to the  component removal ratio, we can utilize stochastic search to find the approximate optimal solution. Second, the model fitting for the spectrum compensation module needs high computing resources; however, the fitting only occurs in the training phase. In the deployment phase, \TN{} can respond fast by selecting the optimal pre-set model. Third, \TN{} performs well in recovering phoneme-level spectrum reduction, but yields nearly 50\% effectiveness against word-level attacks since the word-level attacks obfuscate the phoneme boundaries and cause more shifted phoneme errors. In our future work, we will try to improve the ASR accuracy by recovering the phoneme boundaries of attacked audio.

% However, we can accelerate the DFT computation with specialized DSP chips and reduce the computation time via parallel computing. 
%
%Finally, our system only targets the spectrum reduction attack right now. \sw{However, considering the attack operability, the spectrum reduction attack is riskier than others because other spectrum-based attacks can already be mitigated by effective defenses while the spectrum reduction attack is easier to be implemented compared with complicated adversarial audio attacks.} Also, attackers do not need additional hardware devices, e.g., ultrasonic generator~\cite{Metamorph} or laser emitter~\cite{LightCommands}. In our future work, we tend to build a general acoustic propagation pipeline to defeat more general audio attacks. To achieve this goal, we may integrate more attack detection modules and mitigation modules into the \TN{}.

The \TN{} is evaluated based on word/character recognition, which is widely adopted by most content moderation systems that use keyword filtering~\cite{hettiachchi2019towards} and regular expressions~\cite{iqbal2022looking}. However, some content moderation systems may utilize speech understanding methods that consider the text semantics instead of individual words~\cite{sun2022design}. We are unable to evaluate the \TN{} performance with the speech understanding model due to the lack of a dataset.
Such an evaluation requires a labeled dataset containing problematic content; however, existing public datasets are built for either images or text~\cite{akyon2022deep}, not for audio.

\section{Related Work}

% \vspace{-0.05in}
\subsection{Attacks against ASR Systems}
% \vspace{-0.03in}

\noindent {\bf Attacks against Speaker-Dependent ASRs.}
The speaker-dependent ASRs match the unique voice patterns for specific humans~\cite{OKSiri} and are vulnerable to four types of attacks, i.e., impersonation, replay, speech synthesis, and speech conversion~\cite{SpoofingASR}. Attackers can launch impersonation attacks by physically changing their voices~\cite{IMP1} or launch replay attacks with pre-recorded audio~\cite{replayattack, REPLAY1, REPLAY2}. Attackers can also synthesize the victims' voices by merging acoustic patterns and desired content~\cite{SyntheticSpeech, Synthetic, speech_synthesis_attack}. Moreover, attackers can convert anyone's audio into the victims' styles including timbre~\cite{TimbreConv, TimbreConv2} and prosody~\cite{prosody, Sisman2018}. 

\noindent {\bf Attacks against Speaker-Independent ASRs.}
The speaker-independent ASRs are more vulnerable because they accept voice signals from anyone~\cite{SoK_ASR}. Attackers can use different physical devices to attack the audio capture phase. Dolphin attacks modulate audio into ultrasonic band to exploit the vulnerability of microphone non-linearity~\cite{Dolphin}. Attackers can also modulate malicious audio into laser light without being noticed by humans~\cite{LightCommands}. To leverage the pre-processing filter of ASRs, attackers disguise malicious audio by adding crafted high-frequency noise~\cite{SPHidden}. Moreover, attackers can utilize the psychoacoustics hiding to generate malicious audio below the thresholds of human perception~\cite{Psychoacoustic}.

% By leveraging the principle of psychoacoustics, the feature extraction module can be attacked by time-domain inversion and random phase generation~\cite{SPHidden, Psychoacoustic}.

\noindent {\bf Adversarial Attacks against ASR Inference.}
Attackers can leverage adversarial machine learning to generate malicious audio that can be interpreted by machines but cannot be recognized by humans~\cite{advattackccs21}. The adversarial audios can be noise-like~\cite{HidCmd}, song-like~\cite{CmdSong}, or any format audio~\cite{wang2022ghosttalk}. To directly fool the NLP module after ASRs, skill squatting attacks are proposed to mislead the system and launch malicious applications~\cite{SkillSquatting, DangerousSkills, MIM, AfterASR, dangerousskill}. The above attacks either hide speaker identities or hide voice content from human perception, whereas spectrum reduction attacks~\cite{dft_attack} tend to mislead machines while keeping human perception.

% \vspace{-0.1in}
\subsection{Defenses against Malicious Audio Attacks}
% \vspace{-0.03in}
%
\noindent {\bf Defenses with Frequency and Time Domain Features.}
To detect malicious audio, multiple frequency-based features are commonly used, including Mel-Frequency Cepstral Coefficient (MFCC)~\cite{Xie_2019}, Constant Q Cepstral Coefficients (CQCC)~\cite{CQCC}, and Linear Prediction Cepstral Coefficient (LPCC)~\cite{LPCC, Void}. Despite these cepstral coefficients, malicious audio are usually different over frequency bands and can be detected using high-frequency features~\cite{modreplay}, sub-bass frequency features~\cite{subbass}, and frequency modulation features~\cite{ModDynamic}. Besides the frequency-based features, DualGuard detects malicious audio with both frequency-domain and time-domain features~\cite{modreplay}.

% To detect audio attacks, several audio features are used to distinguish benign and malicious audio, such as Mel-Frequency Cepstral Coefficient (MFCC)~\cite{Xie_2019}, Constant Q Cepstral Coefficients (CQCC)~\cite{CQCC}, and Linear Prediction Cepstral Coefficient (LPCC)~\cite{LPCC, Void}. 

% Other frequency-based features can also be utilized to distinguish malicious audio attacks, such as high-frequency features~\cite{modreplay}, sub-bass frequency features~\cite{subbass}, and frequency modulation features~\cite{ModDynamic}. 

% The malicious audio attacks may also leave clues in the time domain, thus DualGuard tends to combine the time-domain and frequency-domain features together~\cite{modreplay}. 

\noindent {\bf Defenses with Audio Properties.}
Malicious audio attacks can be distinguished by physical properties of audio propagation, e.g., time differences of arrival~\cite{SIEVE, VoiceLive}, Doppler effects~\cite{Liveness}, body-surface vibration~\cite{surfacevibration}, and microphone array fingerprint~\cite{arrayID, RobustDetect}. 
Moreover, the multi-modal method is an alternative to ensure audio integrity~\cite{Multimodal}. Due to the perturbation sensitivity of adversarial audio, denoising algorithms, defensive perturbations, and audible transform are applied to defeat malicious audio~\cite{Dompteur}. WaveGuard identifies malicious audio by analyzing the transcription differences between original and transformed audio~\cite{WaveGuard}. The properties of articulatory phonetics are effective to defeat the adversarial audio deepfakes~\cite{whoareu}. SkillDetective points out potential policy violations to detect skill-related attacks~\cite{SkillDetective}.

% Another idea to distinguish malicious audio determines the audio source by the physical properties of audio propagation, such as body-surface vibration~\cite{surfacevibration}, time differences of arrival~\cite{SIEVE, VoiceLive}, Doppler effects~\cite{Liveness}, and microphone array fingerprint~\cite{arrayID, RobustDetect}. 
% To defeat adversarial audio attacks, researchers can apply denoising algorithms, defensive perturbations, or audible transform~\cite{Dompteur} to the attacked audio.

% WaveGuard detects adversarial audio by incorporating audio transformation functions and analyzes the differences between the transcriptions of original and transformed audio~\cite{WaveGuard}.
% Researchers can also leverage the principle of articulatory phonetics to detect the generative malicious audio \emph{deepfakes}~\cite{whoareu}. To detect skill-related attacks, SkillDetective identifies possible policy violations~\cite{SkillDetective}.

% \vspace{-0.1in}
\section{Conclusion}
% \vspace{-0.05in}

We propose an acoustic compensation system named~\TN{} to mitigate the effects caused by spectrum reduction attacks.
The \TN{} system contains three modules. %, i.e., spectrum compensation module, noise addition module, and adaptation module.
The spectrum compensation module estimates the removed frequency components by using the remaining ones.
The noise addition module models the ambient noise in the acoustic propagation process to add defensive perturbations to the malicious audio.
The adaptation module detects the properties of malicious audio and sets the optimal parameters for each defense module.
Experimental results show both the word error rate and the character error rate decreases significantly when using the~\TN{}.
Among the ASR inference errors caused by spectrum reduction attacks, up to 87.9\% of errors can be eliminated via~\TN{}.
We also conduct an error analysis on the residual ASR inference errors to investigate the root causes and potential mitigation.
%----------------------------------------------------------------
\section*{Acknowledgment}
This work is supported in part by the ONR grant N00014-23-1-2122 and NSFC grant 62132011.
%----------------------------------------------------------------
\bibliographystyle{IEEEtranS}
\bibliography{paperref}

% Generated by IEEEtranS.bst, version: 1.12 (2007/01/11)
\begin{thebibliography}{10}
\providecommand{\url}[1]{#1}
\csname url@samestyle\endcsname
\providecommand{\newblock}{\relax}
\providecommand{\bibinfo}[2]{#2}
\providecommand{\BIBentrySTDinterwordspacing}{\spaceskip=0pt\relax}
\providecommand{\BIBentryALTinterwordstretchfactor}{4}
\providecommand{\BIBentryALTinterwordspacing}{\spaceskip=\fontdimen2\font plus
\BIBentryALTinterwordstretchfactor\fontdimen3\font minus
  \fontdimen4\font\relax}
\providecommand{\BIBforeignlanguage}[2]{{%
\expandafter\ifx\csname l@#1\endcsname\relax
\typeout{** WARNING: IEEEtranS.bst: No hyphenation pattern has been}%
\typeout{** loaded for the language `#1'. Using the pattern for}%
\typeout{** the default language instead.}%
\else
\language=\csname l@#1\endcsname
\fi
#2}}
\providecommand{\BIBdecl}{\relax}
\BIBdecl

\bibitem{SPHidden}
H.~Abdullah, W.~Garcia, C.~Peeters, P.~Traynor, K.~R.~B. Butler, and J.~Wilson,
  ``Practical hidden voice attacks against speech and speaker recognition
  systems,'' in \emph{Proceedings of the 2019 The Network and Distributed
  System Security Symposium (NDSS '19)}, 2019.

\bibitem{abdullah2022attacks}
H.~Abdullah, A.~Karlekar, S.~Prasad, M.~S. Rahman, L.~Blue, L.~A. Bauer,
  V.~Bindschaedler, and P.~Traynor, ``Attacks as defenses: Designing robust
  audio captchas using attacks on automatic speech recognition systems,''
  \emph{arXiv preprint arXiv:2203.05408}, 2022.

\bibitem{dft_attack}
H.~Abdullah, M.~S. Rahman, W.~Garcia, L.~Blue, K.~Warren, A.~S. Yadav,
  T.~Shrimpton, and P.~Traynor, ``{Hear ``no evil'', see ``Kenansville'':
  Efficient and Transferable Black-box Attacks on Speech Recognition and Voice
  Identification Systems},'' in \emph{IEEE Symposium on Security and Privacy
  (IEEE S\&P)}, 2021.

\bibitem{SoK_ASR}
H.~Abdullah, K.~Warren, V.~Bindschaedler, N.~Papernot, and P.~Traynor, ``Sok:
  The faults in our asrs: An overview of attacks against automatic speech
  recognition and speaker identification systems,'' in \emph{2021 IEEE
  Symposium on Security and Privacy (S\&P)}, 2021, pp. 730--747.

\bibitem{Void}
\BIBentryALTinterwordspacing
M.~E. Ahmed, I.-Y. Kwak, J.~H. Huh, I.~Kim, T.~Oh, and H.~Kim, ``Void: A fast
  and light voice liveness detection system,'' in \emph{29th USENIX Security
  Symposium (USENIX Security 20)}.\hskip 1em plus 0.5em minus 0.4em\relax
  USENIX Association, Aug. 2020, pp. 2685--2702. [Online]. Available:
  \url{https://www.usenix.org/conference/usenixsecurity20/presentation/ahmed-muhammad}
\BIBentrySTDinterwordspacing

\bibitem{KWS}
\BIBentryALTinterwordspacing
S.~Ahmed, I.~Shumailov, N.~Papernot, and K.~Fawaz, ``Towards more robust
  keyword spotting for voice assistants,'' in \emph{31st USENIX Security
  Symposium (USENIX Security 22)}.\hskip 1em plus 0.5em minus 0.4em\relax
  Boston, MA: USENIX Association, Aug. 2022, pp. 2655--2672. [Online].
  Available:
  \url{https://www.usenix.org/conference/usenixsecurity22/presentation/ahmed}
\BIBentrySTDinterwordspacing

\bibitem{akyon2022deep}
F.~C. Akyon and A.~Temizel, ``Deep architectures for content moderation and
  movie content rating,'' \emph{arXiv preprint arXiv:2212.04533}, 2022.

\bibitem{wer}
\BIBentryALTinterwordspacing
A.~Ali and S.~Renals, ``Word error rate estimation for speech recognition:
  e-{WER},'' in \emph{Proceedings of the 56th Annual Meeting of the Association
  for Computational Linguistics (Volume 2: Short Papers)}.\hskip 1em plus 0.5em
  minus 0.4em\relax Melbourne, Australia: Association for Computational
  Linguistics, Jul. 2018, pp. 20--24. [Online]. Available:
  \url{https://aclanthology.org/P18-2004}
\BIBentrySTDinterwordspacing

\bibitem{Alexa}
{Amazon Alexa}, ``{Wikipedia}{,} the free encyclopedia,'' 2022,
  \url{https://en.wikipedia.org/wiki/Amazon_Alexa}, [accessed October 2022].

\bibitem{GA}
G.~Assistant, ``{Google},'' 2022, \url{https://assistant.google.com}, [accessed
  October 2022].

\bibitem{subbass}
\BIBentryALTinterwordspacing
L.~Blue, L.~Vargas, and P.~Traynor, ``Hello, is it me you're looking for?:
  Differentiating between human and electronic speakers for voice interface
  security,'' in \emph{Proceedings of the 11th ACM Conference on Security \&
  Privacy in Wireless and Mobile Networks}, ser. WiSec '18.\hskip 1em plus
  0.5em minus 0.4em\relax New York, NY, USA: ACM, 2018, pp. 123--133. [Online].
  Available: \url{http://doi.acm.org/10.1145/3212480.3212505}
\BIBentrySTDinterwordspacing

\bibitem{whoareu}
\BIBentryALTinterwordspacing
L.~Blue, K.~Warren, H.~Abdullah, C.~Gibson, L.~Vargas,
  J.~O{\textquoteright}Dell, K.~Butler, and P.~Traynor, ``Who are you (i really
  wanna know)? detecting audio {DeepFakes} through vocal tract
  reconstruction,'' in \emph{31st USENIX Security Symposium (USENIX Security
  22)}.\hskip 1em plus 0.5em minus 0.4em\relax Boston, MA: USENIX Association,
  Aug. 2022, pp. 2691--2708. [Online]. Available:
  \url{https://www.usenix.org/conference/usenixsecurity22/presentation/blue}
\BIBentrySTDinterwordspacing

\bibitem{Multimodal}
H.~Bredin, A.~Miguel, I.~H. Witten, and G.~Chollet, ``Detecting replay attacks
  in audiovisual identity verification,'' in \emph{2006 IEEE International
  Conference on Acoustics Speech and Signal Processing Proceedings}, vol.~1,
  May 2006, pp. I--I.

\bibitem{bruschi2021low}
V.~Bruschi, S.~Nobili, A.~Terenzi, and S.~Cecchi, ``A low-complexity
  linear-phase graphic audio equalizer based on ifir filters,'' \emph{IEEE
  Signal Processing Letters}, vol.~28, pp. 429--433, 2021.

\bibitem{AdversarySamples}
N.~Carlini and D.~Wagner, ``Audio adversarial examples: Targeted attacks on
  speech-to-text,'' in \emph{2018 IEEE Security and Privacy Workshops (SPW)},
  May 2018, pp. 1--7.

\bibitem{HidCmd}
\BIBentryALTinterwordspacing
N.~Carlini, P.~Mishra, T.~Vaidya, Y.~Zhang, M.~Sherr, C.~Shields, D.~Wagner,
  and W.~Zhou, ``Hidden voice commands,'' in \emph{25th {USENIX} Security
  Symposium ({USENIX} Security 16)}.\hskip 1em plus 0.5em minus 0.4em\relax
  Austin, TX: {USENIX} Association, 2016, pp. 513--530. [Online]. Available:
  \url{https://www.usenix.org/conference/usenixsecurity16/technical-sessions/presentation/carlini}
\BIBentrySTDinterwordspacing

\bibitem{CarrascoMolina2019}
\BIBentryALTinterwordspacing
M.~Carrasco~Molina, \emph{Siri and Search}.\hskip 1em plus 0.5em minus
  0.4em\relax Berkeley, CA: Apress, 2019, pp. 139--188. [Online]. Available:
  \url{https://doi.org/10.1007/978-1-4842-4291-9_7}
\BIBentrySTDinterwordspacing

\bibitem{chen2006new}
J.~Chen, J.~Benesty, Y.~Huang, and S.~Doclo, ``New insights into the noise
  reduction wiener filter,'' \emph{IEEE Transactions on audio, speech, and
  language processing}, vol.~14, no.~4, pp. 1218--1234, 2006.

\bibitem{FakeWake}
\BIBentryALTinterwordspacing
Y.~Chen, Y.~Bai, R.~Mitev, K.~Wang, A.-R. Sadeghi, and W.~Xu, ``Fakewake:
  Understanding and mitigating fake wake-up words of voice assistants,'' in
  \emph{Proceedings of the 2021 ACM SIGSAC Conference on Computer and
  Communications Security}, ser. CCS '21.\hskip 1em plus 0.5em minus
  0.4em\relax New York, NY, USA: Association for Computing Machinery, 2021, p.
  1861–1883. [Online]. Available:
  \url{https://doi.org/10.1145/3460120.3485365}
\BIBentrySTDinterwordspacing

\bibitem{dangerousskill}
\BIBentryALTinterwordspacing
L.~Cheng, C.~Wilson, S.~Liao, J.~Young, D.~Dong, and H.~Hu, ``Dangerous skills
  got certified: Measuring the trustworthiness of skill certification in voice
  personal assistant platforms,'' in \emph{Proceedings of the 2020 ACM SIGSAC
  Conference on Computer and Communications Security}, ser. CCS '20.\hskip 1em
  plus 0.5em minus 0.4em\relax New York, NY, USA: Association for Computing
  Machinery, 2020, p. 1699–1716. [Online]. Available:
  \url{https://doi.org/10.1145/3372297.3423339}
\BIBentrySTDinterwordspacing

\bibitem{CMUSphinx}
CMUSphinx, ``{CMUSphinx: Open Source Speech Recognition Toolkit},'' 2023,
  \url{https://cmusphinx.github.io}, [accessed April 2023].

\bibitem{TIMIT}
L.~D. Consortium, ``{TIMIT} acoustic-phonetic continuous speech corpus.''
  \url{https://catalog.ldc.upenn.edu/LDC93S1}, 2022, accessed September, 2022.

\bibitem{MicroCortana}
M.~Cortana, ``{Microsoft},'' 2022,
  \url{https://www.microsoft.com/en-us/cortana}, [accessed October 2022].

\bibitem{davis1980comparison}
S.~Davis and P.~Mermelstein, ``Comparison of parametric representations for
  monosyllabic word recognition in continuously spoken sentences,'' \emph{IEEE
  transactions on acoustics, speech, and signal processing}, vol.~28, no.~4,
  pp. 357--366, 1980.

\bibitem{SyntheticSpeech}
P.~L. {De Leon}, M.~{Pucher}, J.~{Yamagishi}, I.~{Hernaez}, and I.~{Saratxaga},
  ``Evaluation of speaker verification security and detection of hmm-based
  synthetic speech,'' \emph{IEEE Transactions on Audio, Speech, and Language
  Processing}, vol.~20, no.~8, pp. 2280--2290, Oct 2012.

\bibitem{Dompteur}
\BIBentryALTinterwordspacing
T.~Eisenhofer, L.~Sch{\"o}nherr, J.~Frank, L.~Speckemeier, D.~Kolossa, and
  T.~Holz, ``Dompteur: Taming audio adversarial examples,'' in \emph{30th
  USENIX Security Symposium (USENIX Security 21)}.\hskip 1em plus 0.5em minus
  0.4em\relax USENIX Association, Aug. 2021, pp. 2309--2326. [Online].
  Available:
  \url{https://www.usenix.org/conference/usenixsecurity21/presentation/eisenhofer}
\BIBentrySTDinterwordspacing

\bibitem{SpecPatch}
H.~Guo, Y.~Wang, N.~Ivanov, L.~Xiao, and Q.~Yan, ``Specpatch: Human-in-the-loop
  adversarial audio spectrogram patch attack on speech recognition,'' in
  \emph{Proceedings of the 2022 ACM SIGSAC Conference on Computer and
  Communications Security}, ser. CCS '22, 2022.

\bibitem{deepspeech}
\BIBentryALTinterwordspacing
A.~Y. Hannun, C.~Case, J.~Casper, B.~Catanzaro, G.~Diamos, E.~Elsen,
  R.~Prenger, S.~Satheesh, S.~Sengupta, A.~Coates, and A.~Y. Ng, ``Deep speech:
  Scaling up end-to-end speech recognition,'' \emph{CoRR}, vol. abs/1412.5567,
  2014. [Online]. Available: \url{http://arxiv.org/abs/1412.5567}
\BIBentrySTDinterwordspacing

\bibitem{Impersonation}
R.~G. Hautam{\"a}ki, T.~Kinnunen, V.~Hautam{\"a}ki, T.~Leino, and A.-M.
  Laukkanen, ``I-vectors meet imitators: on vulnerability of speaker
  verification systems against voice mimicry,'' in \emph{INTERSPEECH}, 2013.

\bibitem{OKSiri}
\BIBentryALTinterwordspacing
R.~He, X.~Ji, X.~Li, Y.~Cheng, and W.~Xu, ``"{{{{{OK}}}}}, siri" or "hey,
  google": Evaluating voiceprint distinctiveness via content-based {PROLE}
  score,'' in \emph{31st USENIX Security Symposium (USENIX Security 22)}.\hskip
  1em plus 0.5em minus 0.4em\relax Boston, MA: USENIX Association, Aug. 2022,
  pp. 1131--1148. [Online]. Available:
  \url{https://www.usenix.org/conference/usenixsecurity22/presentation/he-ruiwen}
\BIBentrySTDinterwordspacing

\bibitem{hettiachchi2019towards}
D.~Hettiachchi and J.~Goncalves, ``Towards effective crowd-powered online
  content moderation,'' in \emph{Proceedings of the 31st Australian Conference
  on Human-Computer-Interaction}, 2019, pp. 342--346.

\bibitem{WaveGuard}
\BIBentryALTinterwordspacing
S.~Hussain, P.~Neekhara, S.~Dubnov, J.~McAuley, and F.~Koushanfar,
  ``{WaveGuard}: Understanding and mitigating audio adversarial examples,'' in
  \emph{30th USENIX Security Symposium (USENIX Security 21)}.\hskip 1em plus
  0.5em minus 0.4em\relax USENIX Association, Aug. 2021, pp. 2273--2290.
  [Online]. Available:
  \url{https://www.usenix.org/conference/usenixsecurity21/presentation/hussain}
\BIBentrySTDinterwordspacing

\bibitem{iqbal2022looking}
W.~Iqbal, G.~Tyson, and I.~Castro, ``Looking on efficiency of content
  moderation systems from the lens of reddit's content moderation experience
  during covid-19,'' \emph{Available at SSRN 4007864}, 2022.

\bibitem{alexacontrol}
C.~Jimenez, E.~Saavedra, G.~del Campo, and A.~Santamaria, ``Alexa-based voice
  assistant for smart home applications,'' \emph{IEEE Potentials}, vol.~40,
  no.~4, pp. 31--38, 2021.

\bibitem{REPLAY2}
J.Villalba and E.~Lleida, ``Preventing replay attacks on speaker verification
  systems,'' in \emph{2011 Carnahan Conference on Security Technology}, Oct
  2011, pp. 1--8.

\bibitem{phoneticSegment}
\BIBentryALTinterwordspacing
F.~Kreuk, J.~Keshet, and Y.~Adi, ``Self-supervised contrastive learning for
  unsupervised phoneme segmentation,'' 2020. [Online]. Available:
  \url{https://arxiv.org/abs/2007.13465}
\BIBentrySTDinterwordspacing

\bibitem{SkillSquatting}
\BIBentryALTinterwordspacing
D.~Kumar, R.~Paccagnella, P.~Murley, E.~Hennenfent, J.~Mason, A.~Bates, and
  M.~Bailey, ``Skill squatting attacks on amazon alexa,'' in \emph{Proceedings
  of the 27th USENIX Conference on Security Symposium}, ser. SEC'18.\hskip 1em
  plus 0.5em minus 0.4em\relax Berkeley, CA, USA: USENIX Association, 2018, pp.
  33--47. [Online]. Available:
  \url{http://dl.acm.org/citation.cfm?id=3277203.3277207}
\BIBentrySTDinterwordspacing

\bibitem{Lav2017}
\BIBentryALTinterwordspacing
G.~Lavrentyeva, S.~Novoselov, E.~Malykh, A.~Kozlov, O.~Kudashev, and
  V.~Shchemelinin, ``Audio replay attack detection with deep learning
  frameworks,'' in \emph{Proc. Interspeech 2017}, 2017, pp. 82--86. [Online].
  Available: \url{http://dx.doi.org/10.21437/Interspeech.2017-360}
\BIBentrySTDinterwordspacing

\bibitem{Synthetic}
P.~L.~D. Leon, B.~Stewart, and J.~Yamagishi, ``Synthetic speech discrimination
  using pitch pattern statistics derived from image analysis,'' in
  \emph{INTERSPEECH}, 2012.

\bibitem{li2019adversarial}
J.~Li, S.~Qu, X.~Li, J.~Szurley, J.~Z. Kolter, and F.~Metze, ``Adversarial
  music: Real world audio adversary against wake-word detection system,''
  \emph{Advances in Neural Information Processing Systems}, vol.~32, 2019.

\bibitem{RobustDetect}
\BIBentryALTinterwordspacing
Z.~Li, C.~Shi, T.~Zhang, Y.~Xie, J.~Liu, B.~Yuan, and Y.~Chen, ``Robust
  detection of machine-induced audio attacks in intelligent audio systems with
  microphone array,'' in \emph{Proceedings of the 2021 ACM SIGSAC Conference on
  Computer and Communications Security}, ser. CCS '21.\hskip 1em plus 0.5em
  minus 0.4em\relax New York, NY, USA: Association for Computing Machinery,
  2021, p. 1884–1899. [Online]. Available:
  \url{https://doi.org/10.1145/3460120.3484755}
\BIBentrySTDinterwordspacing

\bibitem{gridsearch}
\BIBentryALTinterwordspacing
P.~Liashchynskyi and P.~Liashchynskyi, ``Grid search, random search, genetic
  algorithm: {A} big comparison for {NAS},'' \emph{CoRR}, vol. abs/1912.06059,
  2019. [Online]. Available: \url{http://arxiv.org/abs/1912.06059}
\BIBentrySTDinterwordspacing

\bibitem{meta_rare}
F.~Lux and N.~T. Vu, ``Meta-learning for improving rare word recognition in
  end-to-end asr,'' in \emph{ICASSP 2021 - 2021 IEEE International Conference
  on Acoustics, Speech and Signal Processing (ICASSP)}, 2021, pp. 5974--5978.

\bibitem{drivernavigate}
A.~Mahr, R.~Serafin, C.~Grajeda, and I.~Baggili, ``Auto-parser: Android auto
  and apple carplay forensics,'' in \emph{Digital Forensics and Cyber Crime},
  P.~Gladyshev, S.~Goel, J.~James, G.~Markowsky, and D.~Johnson, Eds.\hskip 1em
  plus 0.5em minus 0.4em\relax Cham: Springer International Publishing, 2022,
  pp. 52--71.

\bibitem{replayattack}
K.~M. {Malik}, H.~{Malik}, and R.~{Baumann}, ``Towards vulnerability analysis
  of voice-driven interfaces and countermeasures for replay attacks,'' in
  \emph{2019 IEEE Conference on Multimedia Information Processing and Retrieval
  (MIPR)}, March 2019, pp. 523--528.

\bibitem{IMP1}
J.~Mari{\'{e}}thoz and S.~Bengio, ``Can a professional imitator fool a
  gmm-based speaker verification system?'' IDIAP, Idiap-RR Idiap-RR-61-2005,
  2005.

\bibitem{arrayID}
\BIBentryALTinterwordspacing
Y.~Meng, J.~Li, M.~Pillari, A.~Deopujari, L.~Brennan, H.~Shamsie, H.~Zhu, and
  Y.~Tian, ``Your microphone array retains your identity: A robust voice
  liveness detection system for smart speakers,'' in \emph{31st USENIX Security
  Symposium (USENIX Security 22)}.\hskip 1em plus 0.5em minus 0.4em\relax
  Boston, MA: USENIX Association, Aug. 2022, pp. 1077--1094. [Online].
  Available:
  \url{https://www.usenix.org/conference/usenixsecurity22/presentation/meng}
\BIBentrySTDinterwordspacing

\bibitem{TimbreConv}
H.~Ming, D.~Huang, L.~Xie, S.~Zhang, M.~Dong, and H.~Li, ``Exemplar-based
  sparse representation of timbre and prosody for voice conversion,'' in
  \emph{2016 IEEE International Conference on Acoustics, Speech and Signal
  Processing (ICASSP)}, 2016, pp. 5175--5179.

\bibitem{MIM}
\BIBentryALTinterwordspacing
R.~Mitev, M.~Miettinen, and A.-R. Sadeghi, ``Alexa lied to me: Skill-based
  man-in-the-middle attacks on virtual assistants,'' in \emph{Proceedings of
  the 2019 ACM Asia Conference on Computer and Communications Security}, ser.
  Asia CCS '19.\hskip 1em plus 0.5em minus 0.4em\relax New York, NY, USA: ACM,
  2019, pp. 465--478. [Online]. Available:
  \url{http://doi.acm.org/10.1145/3321705.3329842}
\BIBentrySTDinterwordspacing

\bibitem{cm_1}
\BIBentryALTinterwordspacing
P.~Nagarsheth, E.~Khoury, K.~Patil, and M.~Garland, ``Replay attack detection
  using dnn for channel discrimination,'' in \emph{Proc. Interspeech 2017},
  2017, pp. 97--101. [Online]. Available:
  \url{http://dx.doi.org/10.21437/Interspeech.2017-1377}
\BIBentrySTDinterwordspacing

\bibitem{LPCC}
S.~{Novoselov}, A.~{Kozlov}, G.~{Lavrentyeva}, K.~{Simonchik}, and
  V.~{Shchemelinin}, ``Stc anti-spoofing systems for the asvspoof 2015
  challenge,'' in \emph{2016 IEEE International Conference on Acoustics, Speech
  and Signal Processing (ICASSP)}, March 2016, pp. 5475--5479.

\bibitem{VCTK}
\BIBentryALTinterwordspacing
U.~of~Edinburgh. The Centre~for Speech Technology Research~(CSTR), ``Cstr vctk
  corpus: English multi-speaker corpus for cstr voice cloning toolkit (version
  0.92),'' 2023, [accessed 11-January-2023]. [Online]. Available:
  \url{https://datashare.ed.ac.uk/handle/10283/3443}
\BIBentrySTDinterwordspacing

\bibitem{Librispeech}
V.~Panayotov, G.~Chen, D.~Povey, and S.~Khudanpur, ``Librispeech: An asr corpus
  based on public domain audio books,'' in \emph{2015 IEEE International
  Conference on Acoustics, Speech and Signal Processing (ICASSP)}, 2015, pp.
  5206--5210.

\bibitem{backoff_rare}
D.~Pruthi, B.~Dhingra, and Z.~C. Lipton, ``Combating adversarial misspellings
  with robust word recognition,'' in \emph{The 57th Annual Meeting of the
  Association for Computational Linguistics (ACL)}, Florence, Italy, July 2019.

\bibitem{memprofiler}
PyPI, ``memory-profiler package.''
  \url{https://pypi.org/project/memory-profiler/}, 2023, accessed April, 2023.

\bibitem{BackDoor}
\BIBentryALTinterwordspacing
N.~Roy, H.~Hassanieh, and R.~Roy~Choudhury, ``Backdoor: Making microphones hear
  inaudible sounds,'' in \emph{Proceedings of the 15th Annual International
  Conference on Mobile Systems, Applications, and Services}, ser. MobiSys
  '17.\hskip 1em plus 0.5em minus 0.4em\relax New York, NY, USA: ACM, 2017, pp.
  2--14. [Online]. Available: \url{http://doi.acm.org/10.1145/3081333.3081366}
\BIBentrySTDinterwordspacing

\bibitem{InaudibleVC}
\BIBentryALTinterwordspacing
N.~Roy, S.~Shen, H.~Hassanieh, and R.~R. Choudhury, ``Inaudible voice commands:
  The long-range attack and defense,'' in \emph{15th {USENIX} Symposium on
  Networked Systems Design and Implementation ({NSDI} 18)}.\hskip 1em plus
  0.5em minus 0.4em\relax Renton, WA: {USENIX} Association, 2018, pp. 547--560.
  [Online]. Available:
  \url{https://www.usenix.org/conference/nsdi18/presentation/roy}
\BIBentrySTDinterwordspacing

\bibitem{Psychoacoustic}
L.~Sch{\"{o}}nherr, K.~Kohls, S.~Zeiler, T.~Holz, and D.~Kolossa, ``Adversarial
  attacks against automatic speech recognition systems via psychoacoustic
  hiding,'' in \emph{Proceedings of the 2019 The Network and Distributed System
  Security Symposium (NDSS '19)}.\hskip 1em plus 0.5em minus 0.4em\relax
  Internet Society, 2019.

\bibitem{Siri}
A.~Siri, ``{Apple},'' 2022, \url{https://www.apple.com/siri/}, [accessed
  October 2022].

\bibitem{Sisman2018}
\BIBentryALTinterwordspacing
B.~Sisman and H.~Li, ``Wavelet analysis of speaker dependent and independent
  prosody for voice conversion,'' in \emph{Proc. Interspeech 2018}, 2018, pp.
  52--56. [Online]. Available:
  \url{http://dx.doi.org/10.21437/Interspeech.2018-1499}
\BIBentrySTDinterwordspacing

\bibitem{inaudible_0}
\BIBentryALTinterwordspacing
L.~Song and P.~Mittal, ``Poster: Inaudible voice commands,'' in
  \emph{Proceedings of the 2017 ACM SIGSAC Conference on Computer and
  Communications Security}, ser. CCS '17.\hskip 1em plus 0.5em minus
  0.4em\relax New York, NY, USA: Association for Computing Machinery, 2017, p.
  2583–2585. [Online]. Available:
  \url{https://doi.org/10.1145/3133956.3138836}
\BIBentrySTDinterwordspacing

\bibitem{LightCommands}
\BIBentryALTinterwordspacing
T.~Sugawara, B.~Cyr, S.~Rampazzi, D.~Genkin, and K.~Fu, ``Light commands:
  {Laser-Based} audio injection attacks on {Voice-Controllable} systems,'' in
  \emph{29th USENIX Security Symposium (USENIX Security 20)}.\hskip 1em plus
  0.5em minus 0.4em\relax USENIX Association, Aug. 2020, pp. 2631--2648.
  [Online]. Available:
  \url{https://www.usenix.org/conference/usenixsecurity20/presentation/sugawara}
\BIBentrySTDinterwordspacing

\bibitem{sun2022design}
H.~Sun and W.~Ni, ``Design and application of an ai-based text content
  moderation system,'' \emph{Scientific Programming}, vol. 2022, 2022.

\bibitem{ModDynamic}
\BIBentryALTinterwordspacing
G.~Suthokumar, V.~Sethu, C.~Wijenayake, and E.~Ambikairajah, ``Modulation
  dynamic features for the detection of replay attacks,'' in \emph{Proc.
  Interspeech 2018}, 2018, pp. 691--695. [Online]. Available:
  \url{http://dx.doi.org/10.21437/Interspeech.2018-1846}
\BIBentrySTDinterwordspacing

\bibitem{TimbreConv2}
H.~Tang, X.~Zhang, J.~Wang, N.~Cheng, and J.~Xiao, ``Avqvc: One-shot voice
  conversion by vector quantization with applying contrastive learning,'' in
  \emph{ICASSP 2022 - 2022 IEEE International Conference on Acoustics, Speech
  and Signal Processing (ICASSP)}, 2022, pp. 4613--4617.

\bibitem{CQCC}
\BIBentryALTinterwordspacing
M.~Todisco, H.~Delgado, and N.~Evans, ``Constant q cepstral coefficients: A
  spoofing countermeasure for automatic speaker verification,'' \emph{Computer
  Speech \& Language}, vol.~45, pp. 516 -- 535, 2017. [Online]. Available:
  \url{http://www.sciencedirect.com/science/article/pii/S0885230816303114}
\BIBentrySTDinterwordspacing

\bibitem{ModIndex}
J.~Villalba and E.~Lleida, ``Detecting replay attacks from far-field recordings
  on speaker verification systems,'' in \emph{Biometrics and ID Management},
  C.~Vielhauer, J.~Dittmann, A.~Drygajlo, N.~C. Juul, and M.~C. Fairhurst,
  Eds.\hskip 1em plus 0.5em minus 0.4em\relax Berlin, Heidelberg: Springer
  Berlin Heidelberg, 2011, pp. 274--285.

\bibitem{surfacevibration}
\BIBentryALTinterwordspacing
C.~Wang, S.~A. Anand, J.~Liu, P.~Walker, Y.~Chen, and N.~Saxena, ``Defeating
  hidden audio channel attacks on voice assistants via audio-induced surface
  vibrations,'' in \emph{Proceedings of the 35th Annual Computer Security
  Applications Conference}, ser. ACSAC '19.\hskip 1em plus 0.5em minus
  0.4em\relax New York, NY, USA: Association for Computing Machinery, 2019, p.
  42–56. [Online]. Available: \url{https://doi.org/10.1145/3359789.3359830}
\BIBentrySTDinterwordspacing

\bibitem{modreplay}
\BIBentryALTinterwordspacing
S.~Wang, J.~Cao, X.~He, K.~Sun, and Q.~Li, ``When the differences in frequency
  domain are compensated: Understanding and defeating modulated replay attacks
  on automatic speech recognition,'' in \emph{Proceedings of the 2020 ACM
  SIGSAC Conference on Computer and Communications Security}, ser. CCS
  '20.\hskip 1em plus 0.5em minus 0.4em\relax New York, NY, USA: Association
  for Computing Machinery, 2020, p. 1103–1119. [Online]. Available:
  \url{https://doi.org/10.1145/3372297.3417254}
\BIBentrySTDinterwordspacing

\bibitem{SIEVE}
\BIBentryALTinterwordspacing
S.~Wang, J.~Cao, K.~Sun, and Q.~Li, ``{SIEVE}: Secure {In-Vehicle} automatic
  speech recognition systems,'' in \emph{23rd International Symposium on
  Research in Attacks, Intrusions and Defenses (RAID 2020)}.\hskip 1em plus
  0.5em minus 0.4em\relax San Sebastian: USENIX Association, Oct. 2020, pp.
  365--379. [Online]. Available:
  \url{https://www.usenix.org/conference/raid2020/presentation/wang-shu}
\BIBentrySTDinterwordspacing

\bibitem{wang2022ghosttalk}
Y.~Wang, H.~Guo, and Q.~Yan, ``Ghosttalk: Interactive attack on smartphone
  voice system through power line,'' in \emph{Network and Distributed Systems
  Security (NDSS) Symposium}, 2022.

\bibitem{REPLAY1}
Z.~Wang, G.~Wei, and Q.~He, ``Channel pattern noise based playback attack
  detection algorithm for speaker recognition,'' in \emph{2011 International
  Conference on Machine Learning and Cybernetics}, vol.~4, July 2011, pp.
  1708--1713.

\bibitem{speech_synthesis_attack}
\BIBentryALTinterwordspacing
E.~Wenger, M.~Bronckers, C.~Cianfarani, J.~Cryan, A.~Sha, H.~Zheng, and B.~Y.
  Zhao, ``"hello, it's me": Deep learning-based speech synthesis attacks in the
  real world,'' in \emph{Proceedings of the 2021 ACM SIGSAC Conference on
  Computer and Communications Security}, ser. CCS '21.\hskip 1em plus 0.5em
  minus 0.4em\relax New York, NY, USA: Association for Computing Machinery,
  2021, p. 235–251. [Online]. Available:
  \url{https://doi.org/10.1145/3460120.3484742}
\BIBentrySTDinterwordspacing

\bibitem{DFT}
\BIBentryALTinterwordspacing
{Wikipedia contributors}, ``Discrete fourier transform --- {Wikipedia}{,} the
  free encyclopedia,'' 2022, [Online; accessed October 2022]. [Online].
  Available:
  \url{https://en.wikipedia.org/w/index.php?title=Discrete_Fourier_transform&oldid=1102326617}
\BIBentrySTDinterwordspacing

\bibitem{elision}
\BIBentryALTinterwordspacing
{Wikipedia\ contributors}, ``Elision --- {Wikipedia}{,} the free
  encyclopedia,'' 2022, [Online; accessed 29-September-2022]. [Online].
  Available: \url{https://en.wikipedia.org/w/index.php?title=Elision}
\BIBentrySTDinterwordspacing

\bibitem{MFCCfeature}
A.~Winursito, R.~Hidayat, A.~Bejo, and M.~N.~Y. Utomo, ``Feature data reduction
  of mfcc using pca and svd in speech recognition system,'' in \emph{2018
  International Conference on Smart Computing and Electronic Enterprise
  (ICSCEE)}, 2018, pp. 1--6.

\bibitem{SpoofingASR}
\BIBentryALTinterwordspacing
Z.~Wu, N.~Evans, T.~Kinnunen, J.~Yamagishi, F.~Alegre, and H.~Li, ``Spoofing
  and countermeasures for speaker verification,'' \emph{Speech Commun.},
  vol.~66, no.~C, p. 130–153, feb 2015. [Online]. Available:
  \url{https://doi.org/10.1016/j.specom.2014.10.005}
\BIBentrySTDinterwordspacing

\bibitem{Xie_2019}
\BIBentryALTinterwordspacing
Z.~Xie, W.~Zhang, Z.~Chen, and X.~Xu, ``\BIBforeignlanguage{English}{A
  comparison of features for replay attack detection},''
  \emph{\BIBforeignlanguage{English}{Journal of Physics: Conference Series}},
  vol. 1229, no.~1, 05 2019. [Online]. Available:
  \url{http://mutex.gmu.edu/login?url=https://www.proquest.com/scholarly-journals/comparison-features-replay-attack-detection/docview/2566186565/se-2}
\BIBentrySTDinterwordspacing

\bibitem{SkillDetective}
\BIBentryALTinterwordspacing
J.~Young, S.~Liao, L.~Cheng, H.~Hu, and H.~Deng, ``{SkillDetective}: Automated
  {Policy-Violation} detection of voice assistant applications in the wild,''
  in \emph{31st USENIX Security Symposium (USENIX Security 22)}.\hskip 1em plus
  0.5em minus 0.4em\relax Boston, MA: USENIX Association, Aug. 2022, pp.
  1113--1130. [Online]. Available:
  \url{https://www.usenix.org/conference/usenixsecurity22/presentation/young}
\BIBentrySTDinterwordspacing

\bibitem{CmdSong}
\BIBentryALTinterwordspacing
X.~Yuan, Y.~Chen, Y.~Zhao, Y.~Long, X.~Liu, K.~Chen, S.~Zhang, H.~Huang,
  X.~Wang, and C.~A. Gunter, ``Commandersong: A systematic approach for
  practical adversarial voice recognition,'' in \emph{27th {USENIX} Security
  Symposium ({USENIX} Security 18)}.\hskip 1em plus 0.5em minus 0.4em\relax
  Baltimore, MD: {USENIX} Association, 2018, pp. 49--64. [Online]. Available:
  \url{https://www.usenix.org/conference/usenixsecurity18/presentation/yuan-xuejing}
\BIBentrySTDinterwordspacing

\bibitem{Dolphin}
G.~Zhang, C.~Yan, X.~Ji, T.~Zhang, T.~Zhang, and W.~Xu, ``Dolphinattack:
  Inaudible voice commands,'' in \emph{Proceedings of the 2017 ACM SIGSAC
  Conference on Computer and Communications Security}, ser. CCS '17, 2017, pp.
  103--117.

\bibitem{Liveness}
\BIBentryALTinterwordspacing
L.~Zhang, S.~Tan, and J.~Yang, ``Hearing your voice is not enough: An
  articulatory gesture based liveness detection for voice authentication,'' in
  \emph{Proceedings of the 2017 ACM SIGSAC Conference on Computer and
  Communications Security}, ser. CCS '17.\hskip 1em plus 0.5em minus
  0.4em\relax New York, NY, USA: ACM, 2017, pp. 57--71. [Online]. Available:
  \url{http://doi.acm.org/10.1145/3133956.3133962}
\BIBentrySTDinterwordspacing

\bibitem{VoiceLive}
\BIBentryALTinterwordspacing
L.~Zhang, S.~Tan, J.~Yang, and Y.~Chen, ``Voicelive: A phoneme localization
  based liveness detection for voice authentication on smartphones,'' in
  \emph{Proceedings of the 2016 ACM SIGSAC Conference on Computer and
  Communications Security}, ser. CCS '16.\hskip 1em plus 0.5em minus
  0.4em\relax New York, NY, USA: ACM, 2016, pp. 1080--1091. [Online].
  Available: \url{http://doi.acm.org/10.1145/2976749.2978296}
\BIBentrySTDinterwordspacing

\bibitem{DangerousSkills}
N.~Zhang, X.~Mi, X.~Feng, X.~Wang, Y.~Tian, and F.~Qian, ``Dangerous skills:
  Understanding and mitigating security risks of voice-controlled third-party
  functions on virtual personal assistant systems,'' in \emph{IEEE S\&P 2019},
  2019.

\bibitem{AfterASR}
Y.~Zhang, A.~Mendoza, G.~Yang, L.~Xu, P.~Chinprutthiwong, and G.~Gu, ``Life
  after speech recognition: Fuzzing semantic misinterpretation for voice
  assistant applications,'' in \emph{Proceedings of the 2019 The Network and
  Distributed System Security Symposium (NDSS '19)}.\hskip 1em plus 0.5em minus
  0.4em\relax Internet Society, 2019.

\bibitem{advattackccs21}
\BIBentryALTinterwordspacing
B.~Zheng, P.~Jiang, Q.~Wang, Q.~Li, C.~Shen, C.~Wang, Y.~Ge, Q.~Teng, and
  S.~Zhang, ``Black-box adversarial attacks on commercial speech platforms with
  minimal information,'' in \emph{Proceedings of the 2021 ACM SIGSAC Conference
  on Computer and Communications Security}, ser. CCS '21.\hskip 1em plus 0.5em
  minus 0.4em\relax New York, NY, USA: Association for Computing Machinery,
  2021, p. 86–107. [Online]. Available:
  \url{https://doi.org/10.1145/3460120.3485383}
\BIBentrySTDinterwordspacing

\bibitem{prosody}
B.~Şişman, H.~Li, and K.~C. Tan, ``Transformation of prosody in voice
  conversion,'' in \emph{2017 Asia-Pacific Signal and Information Processing
  Association Annual Summit and Conference (APSIPA ASC)}, 2017, pp. 1537--1546.

\end{thebibliography}
%----------------------------------------------------------------
\begin{appendices}
\section{Echo Module}\label{append:multipath}

% % The multipath effect is caused by the sound reflections of objects.
% % We emulate the multipath effect by echo module, which generates the echo by shifting and scaling the original audio signals.
% % Different from the compensation module, the operations in the echo module are performed in the time domain, not in the frequency domain.

When an audio is played in an over-the-air environment, multiple signals from different paths can be received due to the object reflection and audio scattering, i.e., the multipath effect. Since the multipath effect is also an important effect in the audio propagation process, we design an echo module to introduce the multipath perturbations into the attacked audio.

% The echo module aims to emulate the audio multipath effect in the voice propagation process.
% The perturbations introduced in the over-the-air environments mainly include the ambient noise and the reflections of objects.
% While the noising module already emulates the ambient noise, echo module is designed to generate the perturbations caused by audio reflections.
% Because the reflections in real scenarios can lead to multiple arrivals of the same signals and the attenuation of the reflected signals, the echo module needs to \sw{model} these properties of the arrived signal.
% Different from the compensation module, the operations in the echo module are performed in the time domain, not in the frequency domain.

\begin{figure}[t]
    \centering
    \includegraphics[width=0.9\linewidth]{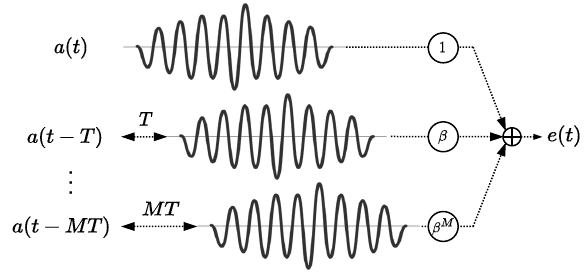}
    % \vspace{-0.05in}
    \caption{The workflow of echo module. The original signal is shifted and attenuated; then, the original signal and all the reflected signals are accumulated as the final received signal.}
    \label{fig:echo_module}
    \vspace{-0.2in}
\end{figure}

% \vspace{0.03in}
{\noindent \bf Modeling.}
We denote the time-domain attacked signal as $a(t)$ and design a basic multipath model to emulate the received signal $e(t)$.
% \vspace{-0.13in}
\begin{equation}
    e(t) = \sum_{i=0}^{M} \mathcal{\beta}^{i} \cdot a(t - i \cdot T),
    % \vspace{-0.08in}
\end{equation}
\noindent where $M$, $\beta$, and $T$ are the parameters of the echo module. $M$ denotes the total times of voice signal reflections; thus, the whole echo signal $e(t)$ contains $M+1$ time-domain waves, including the original voice and $M$ reflected voices. $\beta^{i}$ is the attenuation factor of the $i$-th reflected signal and $T$ is the interval between two consecutive received signals. Therefore, to emulate the multipath effect in the over-the-air propagation, we first shift the original signal in the time domain by $i \cdot T$ ms, where $0\leq i \leq M$. Then, we attenuate the strength of the $i$-th shifted signal with an attenuation factor of  $\beta^{i}$. Finally, we sum up all the sub-signals to get the final echo signal $e(t)$. The multipath process is shown in Figure~\ref{fig:echo_module}. In practice, the echo caused by reflections cannot delay too long and human listeners might not even notice that. Here, we assume the voice is played in an indoor environment and set the longest delay (i.e., $M \cdot T$) to no larger than 15ms (5m / 333m/s).

% \vspace{0.03in}
{\noindent \bf Adaptive Echo.}
Similar to the spectrum compensation module and noise addition module, the best parameters of the echo module are also associated with the component removal ratio. With the increase of the component removal ratio, the attenuation coefficient $\beta$ should become larger to provide more perturbations. Also, the echo module will provide insufficient perturbations if the number of paths ($M$) is too small; however, the echo module can introduce too much interference if it contains too many reflection paths. The shortest time delay $T$ is constrained by $M$ so that $M \cdot T \leq 15$ ms. To obtain the adaptive parameters, we first reproduce the spectrum reduction attacks with different component removal ratios. Then, we find the best parameter combination of echo module for each attack setting. We record the relationships between echo module parameters and component removal ratio. After estimating the component removal ratio of the input voice signal, we load the corresponding best parameters into the echo module to provide adaptive error mitigation against spectrum reduction attacks.

\begin{table}[t]
\begin{center}
\caption{The performance of the echo module on TIMIT against phoneme-level/word-level spectrum reduction attacks.}
% \vspace{-0.1in}
\label{tab:echo}
\renewcommand{\arraystretch}{1}
\resizebox{0.99\linewidth}{!}{
    \begin{tabular}{c|c|c|c|c}
    \toprule
    {\shortstack{\bf Attack \\ \bf Type}} & {\shortstack{\bf Evaluation\\ \bf Metric}} & {\shortstack{\bf Baseline\\ \bf Error}} & {\shortstack{\bf Error \\ \bf w/ Attack}} & {\shortstack{\bf Error \\ \bf w/ Echo}} \\
    \midrule
    \multirow{2}{*}{\shortstack{phoneme-\\level}} & {WER} & {0.217} & {0.597} & {0.537 (-15.8\%)}  \\ 
    {} & {CER} & {0.107} & {0.386} & {0.339 (-16.8\%)}  \\ 
    \midrule
    \multirow{2}{*}{\shortstack{word-\\level}} & {WER} & {0.217} & {0.794} & {0.760 (-5.90\%)}  \\ 
    {} & {CER} & {0.107} & {0.562} & {0.529 (-7.30\%)}  \\ 
    \bottomrule
    \end{tabular}
}
\end{center}
\vspace{-0.15in}
\end{table}

% \vspace{0.03in}
\noindent{\bf Performance of Echo Module.}
We conduct the experiments on the echo module with the optimal parameters. In Table~\ref{tab:echo}, the WER (CER) reduces to 0.537 (0.339) against the phoneme-level attack. Considering the inherent error of 0.217 (0.107), the WER (CER) reduction ratio is 15.8\% (16.8\%). For the word-level attack, the WER (CER) reduces to 0.760 (0.529), where the reduction ratio is 5.9\% (7.3\%) among the errors caused by attacks. The experimental results show the multipath perturbations have limited improvement effects on the ASR inference results compared with spectrum compensation and noise addition modules. The key reason is that the echo module does not introduce components of new frequencies while the other two modules can introduce new frequency components, which target the weakness of spectrum reduction attacks.

\begin{figure}[t]
    \centering
    \subfloat[phoneme-level attack]{
        \includegraphics[width=0.495\linewidth]{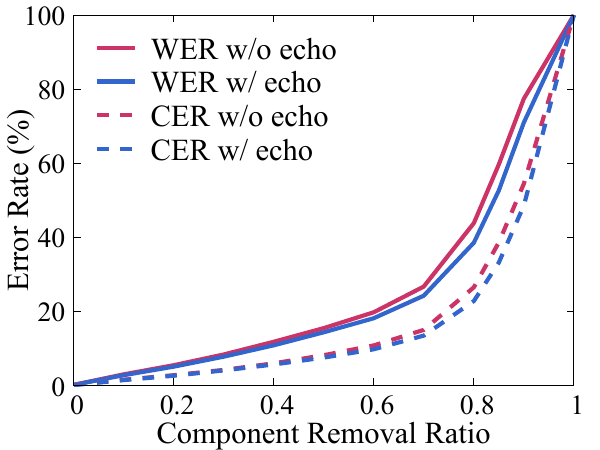}
        }% 
    \subfloat[word-level attack]{
        \includegraphics[width=0.495\linewidth]{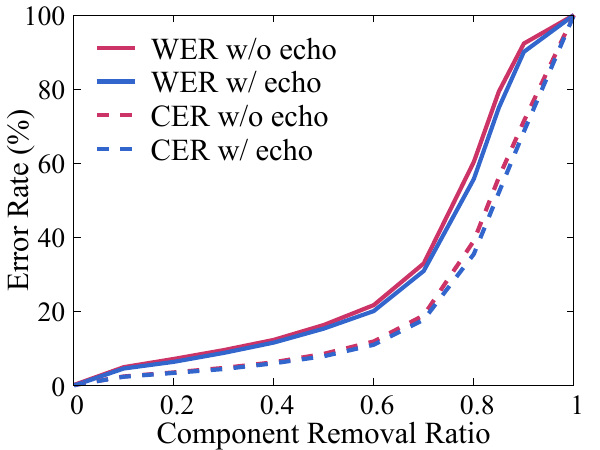}
        }% 
    % \vspace{-0.05in}
    \caption{The echo module performance against word-level and phoneme-level attacks with various component removal ratios.}
    \label{fig:echo_R}
    \vspace{-0.2in}
\end{figure}

\noindent{\bf Impact of Component Removal Ratio.}
We conduct experiments on the echo module over different component removal ratios. The mitigation performance against phoneme-level and word-level attacks is shown in Figure~\ref{fig:echo_R}, where the echo module parameters have adapted to component removal ratios. The mitigation effect against phoneme-level attacks is better than that against word-level attacks, indicating that phoneme-level attacks are more sensitive to time-domain perturbations. Also, the module parameters should change adaptively with the component removal ratios. If we use a set of fixed module parameters, the echo module can reduce the inference errors only if the component removal ratio is larger than a specific threshold. However, if the component removal ratio is smaller than this threshold, the perturbations caused by the echo module might bring negative effects. Therefore, in that case, we need to rely on the adaptation module to adjust the echo module parameters in order to weaken the perturbations, e.g., reduce the attenuation factor or decrease the reflection times.

% \vspace{0.03in}
\noindent{\bf Impact of Echo Module Parameters.}
Different from the spectrum compensation and noise addition modules, the echo module has three interrelated parameters, i.e., the echo interval $T$, the reflection times $M$, and the attenuation factor $\beta$. Hence, it is not practical to use control variable methods due to the high correlations between the module parameters. However, we can set the range for parameters and seek the optimal parameter combination. To achieve this goal, we define some prior rules to narrow down the parameter range. First, the longest distance of delay should not be longer than 5m, i.e., $M\cdot T < 16k \cdot 5 / 333 = 240$. Then, we set the selection interval for each parameter (5 for $T$, 1 for $M$, 0.1 for $\beta$). Finally, given the attack with a specific component removal ratio, we find the optimal parameter combination by the grid search~\cite{gridsearch}. For example, if the component removal ratio is 0.85, the best parameters would be $T=5$, $M=10$, and $\beta=0.4$.

\vspace{-0.05in}
\section{\TN{} Performance tested with CMU Sphinx}\label{append:perf}

\begin{table*}[t]
\begin{center}
\caption{{The performance of \TN{} and its each module against the word-level/phoneme-level spectrum reduction attacks (component removal ratio is 0.85). We evaluate both WER and CER according to the inference results of CMU Sphinx.}}
% \vspace{-0.05in}
\label{tab:append_perf}
\resizebox{.98\textwidth}{!}{
\begin{threeparttable}[b]
    \begin{tabular}{c|c|c|c|c|c|c|c}
    \toprule
    \multirow{2}{*}{\shortstack{\bf Dataset}} & \multirow{2}{*}{\shortstack{\bf Attack\\\bf Granularity}} & \multirow{2}{*}{\shortstack{\bf Evaluation\\\bf Metric$^{\dagger}$}} & \multirow{2}{*}{\shortstack{\bf Baseline\\\bf Error$^{\ddagger}$}} & \multirow{2}{*}{\shortstack{\bf Error w/\\\bf Attack$^{\S}$}} & \multicolumn{3}{c}{\bf Error w/ Our Defense$^{*}$} \\
    \cline{6-8}
    {} & {} & {} & {} & {} & {\bf Compensation} & {\bf Noise Addition} & {\bf \TN{}} \\
    \midrule
    % TIMIT
    \multirow{4}{*}{TIMIT} & \multirow{2}{*}{\shortstack{phoneme-\\level}} & {WER} & {0.410} & {0.927} & {0.574 (-68.3\%)} & {0.575 (-68.1\%)} & {0.568 (-69.4\%)} \\ 
    {} & {} & {CER} & {0.223} & {0.656} & {0.391 (-61.2\%)} & {0.387 (-62.1\%)} & {0.382 (-63.3\%)} \\ 
    \cmidrule{2-8}
    {} & \multirow{2}{*}{\shortstack{word-\\level}} & {WER} & {0.410} & {0.937} & {0.767 (-32.3\%)} & {0.757 (-34.2\%)} & {0.750 (-35.5\%)} \\ 
    {} & {} & {CER} & {0.223} & {0.655} & {0.505 (-34.7\%)} & {0.494 (-37.3\%)} & {0.486 (-39.1\%)} \\ 
    \midrule
    % VCTK
    \multirow{4}{*}{VCTK} & \multirow{2}{*}{\shortstack{phoneme-\\level}} & {WER} & {0.575} & {0.993} & {0.702 (-69.6\%)} & {0.716 (-66.3\%)} & {0.688 (-73.0\%)} \\
    {} & {} & {CER} & {0.336} & {0.784} & {0.450 (-74.6\%)} & {0.465 (-71.2\%)} & {0.428 (-79.5\%)} \\
    \cmidrule{2-8}
    {} & \multirow{2}{*}{\shortstack{word-\\level}} & {WER} & {0.575} & {0.969} & {0.819 (-38.1\%)} & {0.823 (-37.1\%)} & {0.802 (-42.4\%)} \\
    {} & {} & {CER} & {0.336} & {0.734} & {0.558 (-44.2\%)} & {0.565 (-42.5\%)} & {0.545 (-47.5\%)} \\
    
    \bottomrule
    \end{tabular}

    \begin{tablenotes}
        \item[$\dagger$] WER: word error rate between labels and predictions; CER: character error rate between labels and predictions.
        \item[$\ddagger$] Baseline Error indicates the average error rate when ASR infers original benign audio.  
        \item[$\S$] Error w/ Attack indicates the average error rate under spectrum reduction attack (including the baseline error).
        \item[$\star$] The percentage in parenthesis represents the reduction ratio to the errors caused by attacks.
    \end{tablenotes}

\end{threeparttable}
}
\end{center}
\vspace{-0.2in}
\end{table*}

\begin{table*}[t]
\begin{center}
\caption{{Performance comparison of \TN{} against spectrum reduction attack variants with different ratios of the retained energy (tested on the VCTK dataset and evaluated using DeepSpeech).}}
% \vspace{-0.05in}
\label{tab:append_energy_25}
\resizebox{0.6\linewidth}{!}{
\begin{threeparttable}[b]
    \begin{tabular}{c|c|c|c|c}
    \toprule
    \multirow{2}{*}{\shortstack{\bf Energy\\\bf Retained}} & \multicolumn{2}{c|}{\bf WER} & \multicolumn{2}{c}{\bf CER} \\
    \cmidrule{2-5}
    {} & {\bf w/ Attack} & {\bf w/ \TN{}} & {\bf w/ Attack} & {\bf w/ \TN{}}\\
    \midrule
    % VCTK
    {0\%$^\dagger$} & {0.885} & {0.686 (-50.0\%)} & {0.688} & {0.506 (-58.1\%)} \\
    % \midrule
    {25\%} & {0.781} & {0.644 (-46.6\%)} & {0.582} & {0.465 (-56.5\%)} \\
    {50\%} & {0.662} & {0.581 (-46.3\%)} & {0.461} & {0.412 (-57.0\%)} \\
    {75\%} & {0.523} & {0.503 (-44.4\%)} & {0.397} & {0.383 (-63.6\%)} \\
    \midrule
    {100\%$^\star$} & \multicolumn{2}{c|}{0.487} & \multicolumn{2}{c}{0.375} \\
    \bottomrule
    \end{tabular}

    \begin{tablenotes}
        \item[$\dagger$] Retained 0\% energy means using the regular spectrum reduction attacks.
        \item[$\star$] Retained 100\% energy means no attack and the errors are baseline errors.
    \end{tablenotes}

\end{threeparttable}
}
\end{center}
\vspace{-0.15in}
\end{table*}

We test the \TN{} performance using another ASR model, i.e., CMU Sphinx, which is a state-of-the-art efficient speech recognition system designed specifically for low-resource platforms. We evaluate the error rates (i.e., WER and CER) by transcribing the audio signals by CMU Sphinx. Table~\ref{tab:append_perf} shows the experimental results with/without the \TN{} defense system.

Basically, the trends of mitigation effects are consistent with those demonstrated in Table~\ref{tab:all} that tests the performance with DeepSpeech. When tested on the TIMIT dataset, the original WER and CER is 0.410 and 0.223, respectively, without any attack and defense. With the spectrum reduction attacks, the WER (CER) increases to 0.927 (0.656) for phoneme-level attacks and increases to 0.937 (0.655) for word-level attacks. This means more than twice the errors occur when the CMU Sphinx suffers from spectrum reduction attacks. However, with the mitigation from the \TN{} system, the WER (CER) error rate reduces by 69.4\% (63.3\%) against phoneme-level attacks and 35.5\% (39.1\%) against word-level attacks. Similar to the results in Table~\ref{tab:all}, the mitigation effects of \TN{} is better than any individual defense module, i.e., spectrum compensation module or noise addition module. Also, the error reduction ratio against phoneme-level attacks is better than that against word-level attacks. All of these patterns are also applicable to the test results on the VCTK dataset.

There are still some differences between the evaluation results using DeepSpeech and CMU Sphinx.
First, with the CMU Sphinx model, the baseline errors are basically more than those tested by DeepSpeech. This is because the CMU Sphinx is prone to misinterpreting audio due to its design for low-resource platforms with a small vocabulary and lightweight language model.
Second, in Table~\ref{tab:append_perf}, the WERs are over 0.9 when suffering from the spectrum reduction attacks, which means the CMU Sphinx is more vulnerable to the audio attacks compared with the DeepSpeech model. 
However, by utilizing the \TN{} system, the WERs can even drop to as low as 0.568.

Overall, both evaluation tests prove the \TN{} system can mitigate the effects caused by the spectrum reduction attacks.

\section{\TN{} Performance against Attack Variants}\label{append:half_energy}

We conduct experiments to consider the situation where attackers attenuate the energy of frequency components to a low level instead of to zeros. In the experiments, word-level attacks are conducted on the VCTK dataset and both the attacked/mitigated audios are evaluated using the DeepSpeech model. We select the most typical experimental settings, i.e., reduce the energy to 25\%, 50\%, and 75\%, respectively.

From the experimental results in Table~\ref{tab:append_energy_25}, we can find two conclusions. First, this attack method will naturally impair the attack performance because only removing part of the energy may not sufficiently change the statistical distribution of audio spectrum. Therefore, in Table~\ref{tab:append_energy_25}, the error rates reduce significantly when attackers retain more energy for the weak components. Second, our defense can be still applied to this attack variant. Though this attack causes a smaller estimated $R$ value hence reducing the actual compensated energy; however, the compensated energy is enough because there is still part of the energy retained, which can also be seen as a part of compensation. For example, the CER reduction rate is 57.0\% for the word-level attack variant, compared to 58.1\% for the regular spectrum reduction attack.

\end{appendices}

% \input{summary}
%----------------------------------------------------------------
\end{document}